\documentclass[aps,tightenlines,twocolumn,floats,prd,superscriptaddress,showpacs,floatfix]{revtex4-1}
\usepackage{graphicx}% Include Fig. files
\usepackage{epsf} 
\usepackage{amsfonts}
\usepackage{amsmath}
\usepackage{amssymb}
\usepackage{bbold}
\usepackage{bm}
\usepackage{bbm}
\usepackage{slashed}
\usepackage{dcolumn}
\usepackage[dvipsnames]{xcolor}
\usepackage{float}
\usepackage[caption=false]{subfig} % <=== "caption=false" added
\usepackage{lipsum}% just to generate text for the example

\newcommand{\tilp}{{\tilde p}}
\newcommand{\tilk}{{\tilde k}}
\newcommand{\cp}{{\bm \sigma}\cdot{\hat{\bf p}}}
\newcommand{\ck}{{\bm \sigma}\cdot{\hat{\bf k}}}

\usepackage{booktabs,siunitx}

\begin{document} 

\title{Covariant spectator theory of quark-antiquark bound states: \\Mass spectra and vertex functions of heavy and heavy-light mesons}

\author{Sofia Leit\~ao}
 \email{sofia.leitao@tecnico.ulisboa.pt}
\affiliation{Centro de F\' isica Te\' orica de Part\' iculas, Instituto Superior T\'ecnico, Av.\ Rovisco Pais, 
1049-001 Lisboa, Portugal}

\author{Alfred Stadler}
\email{stadler@uevora.pt}
\affiliation{Departamento de F\'isica, Universidade de \'Evora, 7000-671 \'Evora, Portugal}
\affiliation{Centro de F\' isica Te\' orica de Part\' iculas, Instituto Superior T\'ecnico, Av.\ Rovisco Pais, 
1049-001 Lisboa, Portugal}

 \author{M. T. Pe\~na}
 \email{teresa.pena@tecnico.ulisboa.pt}
\affiliation{Departamento de F\'isica, Instituto Superior T\'ecnico, Av.\ Rovisco Pais, 
1049-001 Lisboa, Portugal}
\affiliation{Centro de F\' isica Te\' orica de Part\' iculas, Instituto Superior T\'ecnico, Av.\ Rovisco Pais, 
1049-001 Lisboa, Portugal}
 
 \author{Elmar P. Biernat}
 \email{elmar.biernat@tecnico.ulisboa.pt}
 \affiliation{Centro de F\' isica Te\' orica de Part\' iculas, Instituto Superior T\'ecnico, Av.\ Rovisco Pais, 
1049-001 Lisboa, Portugal}

\date{\today}


\begin{abstract}
We use the covariant spectator theory with an effective quark-antiquark interaction, containing Lorentz scalar, pseudoscalar, and vector contributions, to calculate the masses and vertex functions of, simultaneously, heavy and heavy-light mesons. 
We perform least-square fits of the model parameters, including the quark masses, to the meson spectrum and systematically study the sensitivity of the parameters with respect to different sets of fitted data. We investigate the influence of the vector confining interaction by using a continuous parameter controlling its weight. 
We find that vector contributions to the confining interaction between 0 \% and about 30 \% lead to essentially the same agreement with the data. Similarly, the light quark masses are not very tightly constrained.
In all cases, the meson mass spectra calculated with our fitted models agree very well with the experimental data. 
We also calculate the mesons wave functions in a partial wave representation and show how they are related to the meson vertex functions in covariant form.   
\end{abstract}

\maketitle

\section{Introduction}
\label{sec:intro}

A complete and detailed explanation of the meson spectrum from QCD is still lacking. 
Fortunately, with the strong activity at various experimental facilities (LHCb, {\it BABAR}, BES, Belle), and even more high-accuracy experiments scheduled to come online in the near future (GlueX, SuperKEKB, PANDA), a steadily increasing wealth of data on known and newly discovered meson states is now available, and should help us to improve our understanding of these systems.

On the theoretical side, QCD calculations on the lattice are speedily progressing with respect to managing finite volume effects and decreasing pion mass (e.g. \cite{McNeile:2006,Burch:2006,Wada:2007,Dudek:2008,Gregory:2012}, and references therein). For comprehensive reviews on the subject see \cite{Brambilla2014,Klempt20071}.

In parallel to lattice calculations, a variety of non-perturbative continuum approaches have provided important information on the inner workings of mesons. They include nonrelativistic effective field theories for heavy quarkonia \cite{Brambilla:2004, RevModPhys.77.1423}, the Dyson-Schwinger-Bethe-Salpeter (DS-BS) framework \cite{PhysRevC.55.2649,Maris:1997, Ivanov:1998, Maris:1999dk,Maris:2000,Holl:2004,Maris:2005,Holl:2005wq,Maris:2006,Bhagwat:2006,PhysRevD.67.094020, PhysRevC.77.042202, PhysRevC.79.012202, Krassnigg:2010,Blank:2011,Popovici:2014,Hilger:2015ty,Eichmann:2016bf}, which takes dynamical momentum-dependent quark masses into account and is successful in particular in light quark systems, covariant two-body Dirac equations \cite{Crater:2010pi}, two-fermion calculations in relativistic quantum mechanics \cite{Giachetti:2013uo}, and the basis light-front quantization approach \cite{Li:2015zda,Li:2017mlw} with an effective confining Hamiltonian from light-front holographic QCD, which was applied in studies of heavy quarkonia.

Our work uses the Covariant Spectator Theory (CST) \cite{Gross:1969eb,Gross:1982, Gross:1991te, Savkli:2001os, Stadler:2011to, Biernat:2014jt}. This framework belongs to a class of three-dimensional ``quasi-potential'' equations which are derived from the Bethe-Salpeter equation (BSE) by placing constraints on the relative-energy component of a two-particle system.

The CST  framework has attractive features that are worth enumerating here:
(i) It is manifestly covariant, which allows an exact calculation of boosts of two-particle amplitudes.
(ii) It possesses the correct one-body limit, i.e., it turns into an effective one-body Dirac or Klein-Gordon equation when one of the two constituent particles becomes infinitely heavy.
(iii) It has a smooth nonrelativistic limit, in which it reduces to the Schr\"odinger equation.
(iv) It defines ``relativistic wave functions'' which become proper nonrelativistic wave functions in the nonrelativistic limit. One can identify wave function components of purely relativistic origin and get a direct, intuitive picture of the importance of relativity in different systems.
(v) It implements dynamical chiral symmetry breaking, satisfying the axialvector Ward-Takahashi identity. This key feature was absent from previous calculations of quark-antiquark bound states with other 3D reductions of the BS equation \cite{Koll:2000,Spence:1993bh,TIEMEIJER199238}, as well as from the well-known ``relativized''  calculations with Cornell-type potentials \cite{Godfrey:1985}. The implementation of chiral symmetry constraints in CST calculations through a Nambu--Jona-Lasinio mechanism was introduced in \cite{Gross:1991te,GMilana:1994,Savkli:2001os}, and extended more recently in \cite{Biernat:2014jt,Biernat:2014uq}.
(vi) The CST two-quark kernel, determined in the two-body bound-state problem, can later be included consistently in Faddeev-type three-quark calculations of baryons by boosting two-quark rest-frame amplitudes appropriately. Although genuine three-body calculations for baryons in CST have not yet been carried out, the same principle applies to two- and three-nucleon systems where CST has been used extensively and with remarkable success \cite{Sta97,Gro08,Gro10}.

CST is, in some aspects, close to the DS-BS approach, in the sense that both aim at a unified, self-consistent quantum-field-theoretical description of hadrons. But there are also significant differences: DS-BS is formulated in Euclidean space, whereas CST works in Minkowski space. DS-BS implements confinement through the absence of real mass poles of the quark propagators, whereas in CST confinement is the consequence of a confining interaction kernel.

Heavy and heavy-light mesons are very suitable systems to test different mechanisms of confinement, and to possibly determine its Lorentz structure. A confining interaction increases in strength with the distance between quarks, and in higher excited states it should therefore become more important than the short-range one-gluon-exchange (OGE) interaction. The vector meson bottomonium spectrum is particularly interesting in this regard because of the exceptionally high number of excited states below the open-flavor threshold that have already been measured. So far, lattice QCD and DS-BS calculations are having difficulties describing higher excited states \cite{Meinel:2010pv,Meinel:2009rd,Lewis:2012ir,Gray:2005ur,Eichten:1978tg,Dudek:2007wv,Dowdall:2011wh,Burch:2009az,Brambilla:2010cs,Fischer:2015lq}.

In \cite{Leitao:2017it}, we reported on first results of CST calculations of the heavy and heavy-light meson spectrum. We found that a remarkably good description of the masses of mesons with at least one charm or bottom quark can be obtained with a simple covariant interaction kernel, which was chosen to reduce to a Cornell-type potential in the nonrelativistic limit. Only the three strength parameters for a (Lorentz scalar and pseudoscalar) linear confinement, a OGE, and a constant interaction were adjusted in the fits to the data, whereas quark masses and a Pauli-Villars regularization mass were fixed ad-hoc at reasonable values. What is particularly interesting about the results is that we performed \emph{global} least-square fits, such that the three parameters are the same in all sectors when we calculate the whole spectrum, ranging from the $D$ mesons with masses below 2 GeV up to bottomonium with masses above 10 GeV. 

In this work we go beyond \cite{Leitao:2017it} in several aspects.  In addition to the previously used scalar+pseudoscalar Lorentz structure, we introduce a vector interaction, whose relative weight can be altered through a continuous mixing parameter $y$. This is done in a way that in the nonrelativistic limit always the same linear potential is obtained. By letting the parameter $y$ be determined through a fit, we can investigate to what extent the mass spectrum of heavy and heavy-light mesons constrains the Lorentz structure of the confining interaction. 

We also devised a numerical method that makes it feasible to treat the quark masses as adjustable fit parameters. Not only is it interesting to find out how much these masses are constrained by the data, but also how much improvement one can obtain in the quality of the fits when more adjustable parameters are introduced. 

Another interesting question is how sensitive the results are with respect to the selection of the used experimental data. In \cite{Leitao:2017it} we found that fits to a small number of pseudoscalar states alone already yield a model that predicts all other considered mesons with $J\le 1$ with almost the same accuracy as more general fits, indicating that the covariance of the kernel correctly determines the spin-dependence of the interaction. 

The CST wave functions are then analyzed in detail. This provides a means to determine its spin and orbital angular momentum content, which is very useful for the identification of each calculated state. We also examine the wave functions of excited states in dependence of the excitation level, and the size of wave function components of relativistic origin with different quark masses.

In addition to these numerical results, we also present details of the formalism, in particular the form of the CST equations for the general case of unequal masses, the reduction of the one-channel CST equation to partial-wave form, and the relation between the radial wave functions and the covariant form of the corresponding meson vertex function.

This paper is organized  as follows: in Sec.~\ref{sec:formalism} we derive the CST equations and two of its approximations,  one of which is then used in the numerical calculations presented and discussed in Sec.~\ref{sec:results}. In Sec.~\ref{sec:conclusions} we summarize and present our conclusions.

\section{Formalism}
\label{sec:formalism}

\subsection{The four-channel CST equation}
The four-channel CST equation for bound-states of equal mass quarks and antiquarks has been introduced in Refs.~\cite{Savkli:2001os,Biernat:2014jt}. In this work we are interested in cases with unequal masses as well, so we have to generalize the CST equation accordingly.

\begin{figure}[tb]
\begin{center}
 \includegraphics[width=.3\textwidth]{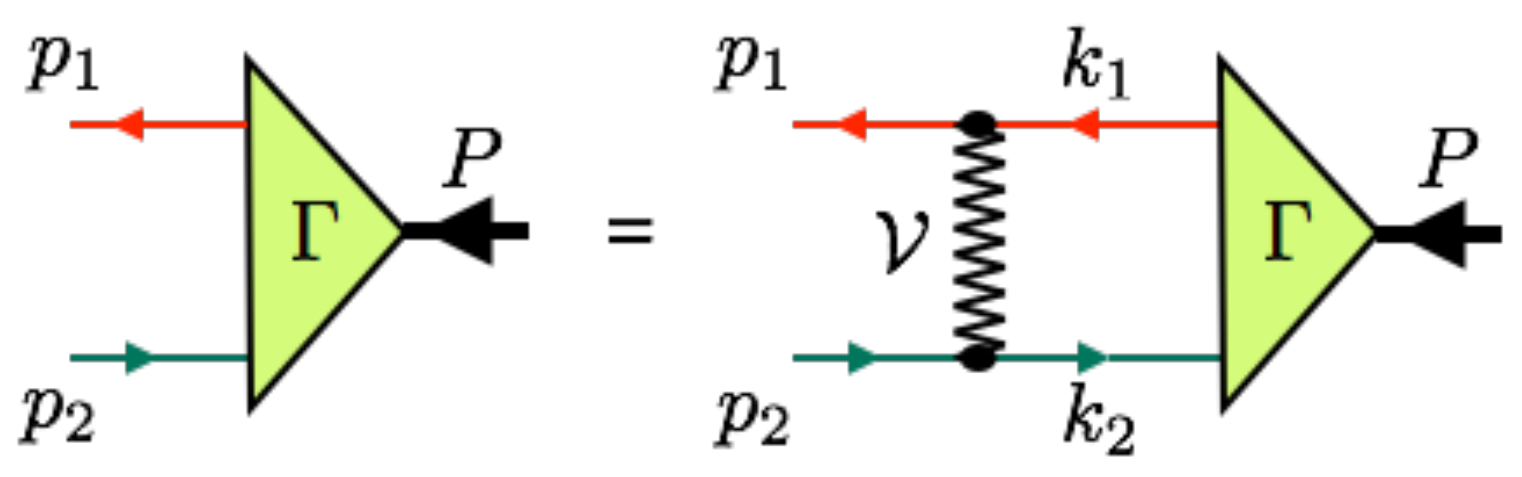}
\caption{Graphic representation of the BSE for the $q\bar{q}$ bound state vertex function $\Gamma$, where ${\cal V}$ represents the kernel of two-body irreducible Feynman diagrams.}
\label{fig:BSE}
\end{center}
\end{figure}

The CST equation can be derived from the BSE for the vertex function $\Gamma_\mathrm{BS}$  (also shown graphically in Fig.~\ref{fig:BSE}),
\begin{multline}
\Gamma_\mathrm{BS}(p_1,p_2) = i \int\frac{d^4k}{(2\pi)^4}\,{\cal V}(p,k;P) 
\\
\times S_1({k}_1)\, \Gamma_\mathrm{BS}(k_1,k_2)\,  S_2(k_2)\, ,
\label{eq:BSE}
\end{multline}
where $S_i(k_i)$ is the dressed propagator of quark $i$ (with an imaginary factor $(-i)$ removed), $P=p_1-p_2$ the total four-momentum, and $p=\frac 12 (p_1+p_2)$ is the relative momentum. The individual quark momenta  $p_i$ in terms of the relative and total momentum are $p_1=p+P/2$ and $p_2=p-P/2$. Analogous expressions relate the intermediate individual quark momenta $k_i$  to the intermediate relative momentum $k$ and to the total momentum.

 The kernel is of the form 
\begin{equation}
{\cal V}(p,k;P) = \frac {3}{4}{\bf F}_1 \cdot  {\bf F}_2 \sum_K V^K (p,k;P) {\Theta}^K_1 \otimes  {\Theta}^K_2 \, ,
\label{eq:BSkernelV}
\end{equation}
where ${\Theta}^K_1$ and ${\Theta}^K_2$ are Dirac matrices, whose type is labeled $K$, associated with the vertices involving quark 1 or 2, respectively. We use ${\Theta}^s_i={\bf 1}$ for scalar, $\Theta^p_i=\gamma^5$ for pseudoscalar, and $\Theta^v_i=\gamma^\mu$ for vector coupling (the Lorentz vector index $\mu$ carried by $\Theta^v_i$  is not explicitly shown when we refer to ${\Theta}^K_i$ in general). The $V^K (p,k;P)$ are covariant scalar functions describing the corresponding momentum dependence. 
However, the explicit dependence of the kernel ${\cal V}$ and the functions $V^K$ on the total momentum $P$ will be suppressed from here on.  The color SU(3) generators, in terms of the Gell-Mann matrices, are  $F_a=\frac12\lambda_a$. All calculations of this paper are performed for color singlet states, for which the color factor becomes $\frac 34 \langle {\bf F}_1 \cdot  {\bf F}_2 \rangle = 1$. 

Note that the multiplication with the kernel in (\ref{eq:BSE}) is an abbreviation that should be interpreted as
\begin{multline}
{\cal V}(p,k) 
 S_1({k}_1)\, \Gamma_\mathrm{BS}(k_1,k_2)\,  S_2(k_2) \equiv\\
 \sum_K V^K (p,k) {\Theta}^K_1  S_1({k}_1)\, \Gamma_\mathrm{BS}(k_1,k_2)\,  S_2(k_2) {\Theta}^K_2 \, .
\end{multline}

In this work we do not calculate the quark self-energies and dynamical masses, but assume constant quark masses $m_i$ instead. The propagators are then
\begin{equation}
S_i(k) = \frac{m_i+\slashed{k}}{m_i^2-k^2-i\epsilon} \, .
\end{equation}

The CST equation is obtained by performing the integration over the energy component of the loop four-momentum, but keeping only the contributions from the poles of the quark propagators. The rationale for discarding the poles in the kernel is mainly that the residues of ladder and crossed-ladder diagrams tend to cancel, in all orders of the coupling constant, in particular when one of the two quark masses becomes large \cite{Gross:1969eb,Gross:1982,Stadler:2011to}. Details about how this integration is evaluated are given in \cite{Biernat:2014jt}. The only difference to \cite{Biernat:2014jt} is that here we have to keep $S_1$ and $S_2$ distinct because of the difference in the quark masses. 

In the following we work in the rest frame of the meson, where $P=(\mu,\bf{0})$, and the quark three-momenta and the relative three-momentum are equal, ${\bf k}_1={\bf k}_2={\bf k}$. We also define $E_{ik} \equiv (m_i^2+{\bf k}^2)^{1/2}$,  the  four-momentum $\hat{k}_i^\pm \equiv (\pm E_{ik},{\bf k})$ of a quark on its positive- or negative-energy mass shell, and the corresponding positive- or negative-energy projector $\Lambda_i(\hat{k}_i^\pm)=(m_i+\slashed{\hat{k}}_i^\pm)/2m_i$.

Closing the $k_0$ integration contour in the lower half plane and keeping only the residues from the quark propagator poles yields
\begin{align}
&  \Gamma_\mathrm{lower}(p_1,p_2)  = \nonumber \\
& - \int_{{\bf k}_1} \mathcal{V}(p,\hat{k}_1^+-P/2)\Lambda_1(\hat{k}_1^+) \Gamma(\hat{k}_1^+,\hat{k}_1^+-P) S_2(\hat{k}_1^+-P) 
\nonumber  \\
& - \int_{{\bf k}_2}  \mathcal{V}(p,\hat{k}_2^++P/2) S_1(\hat{k}_2^+ +P) \Gamma(\hat{k}_2^+ +P,\hat{k}_2^+) \Lambda_2(\hat{k}_2^+)
 \,,
 \label{eq:4CSTlower}
\end{align}
whereas closing it in the upper half plane gives
\begin{align}
&  \Gamma_\mathrm{upper}(p_1,p_2)  = \nonumber \\
& - \int_{{\bf k}_1} \mathcal{V}(p,\hat{k}_1^- -P/2)\Lambda_1(\hat{k}_1^-) \Gamma(\hat{k}_1^-,\hat{k}_1^- -P) S_2(\hat{k}_1^- -P) 
\nonumber  \\
& - \int_{{\bf k}_2}  \mathcal{V}(p,\hat{k}_2^- +P/2) S_1(\hat{k}_2^- +P) \Gamma(\hat{k}_2^- +P,\hat{k}_2^-) \Lambda_2(\hat{k}_2^-)
 \,,
 \label{eq:4CSTupper}
\end{align}
where we have introduced the convenient shorthand
\begin{equation}
\int _{{\bf k}_i} \equiv \int \frac{\mathrm d^3 k}{(2\pi)^3}\frac{m_i}{E_{ik}}
\end{equation}
for the covariant integration measure.

$ \Gamma_\mathrm{lower}(p_1,p_2)$ and $ \Gamma_\mathrm{upper}(p_1,p_2)$ are not necessarily equal, because only the residues of the quark propagator poles were taken into account. The CST vertex function is defined as the symmetric combination
\begin{equation}
 \Gamma(p_1,p_2) \equiv \frac 12 \left[ \Gamma_\mathrm{lower}(p_1,p_2) + \Gamma_\mathrm{upper}(p_1,p_2) \right] \, .
 \label{eq:CSTvfsym}
\end{equation}
In the equal-mass case, the charge-conjugation symmetry of the BSE is preserved when this symmetrized combination of lower and upper half-plane contour integration is used \cite{Biernat:2014jt,Savkli:2001os}.

Before writing the equation for the  CST vertex function (\ref{eq:CSTvfsym}), it is convenient to simplify our notation by  expressing the negative-energy on-shell momenta $\hat{k}_i^-$ in (\ref{eq:4CSTupper}) in terms of the positive-energy on-shell momenta $\hat{k}_i^+$: inverting the integration three-momentum ${\bf k} \rightarrow -{\bf k}$ permits us to write $\hat{k}_i^- \rightarrow - \hat{k}_i^+$. Now we can drop the superscript $\pm$ with the understanding that all on-shell momenta are on the positive-energy mass shell, i.e.\ $\hat{k}_i \equiv \hat{k}_i^+$.

\begin{figure}[tb]
\begin{center}
  \includegraphics[width=.48\textwidth]{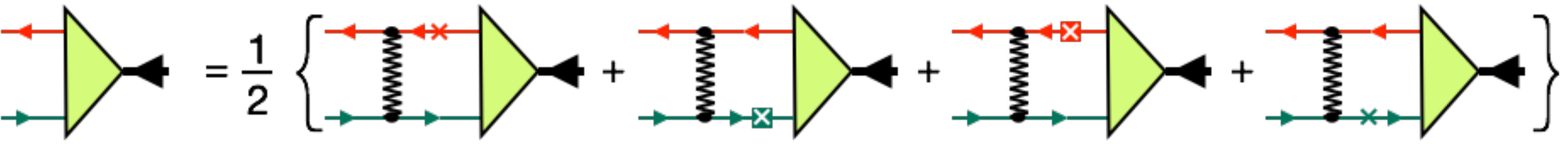}
  \caption{The BS vertex function approximated as a sum of CST vertex functions (crosses on quark lines indicate that a positive-energy pole of the propagator is calculated, light crosses in a dark square refer to a negative-energy pole).}
\label{fig:CST-BS}
\end{center}
\end{figure}

With this notation, the symmetrized CST vertex function is 
\begin{align}
& \Gamma(p_1,p_2) =
-\frac 12 \sum_{\eta=\pm}  \Biggl[
 \int_{{\bf k}_1} 
 \mathcal{V}(p,\eta\hat{k}_1-P/2)\Lambda_1(\eta\hat{k}_1)   
\nonumber \\
& \times \Gamma(\eta\hat{k}_1,\eta\hat{k}_1-P) S_2(\eta\hat{k}_1-P) 
+  \int_{{\bf k}_2} 
 \mathcal{V}(p,\eta\hat{k}_2+P/2) 
 \nonumber \\
 & 
\times S_1(\eta\hat{k}_2+P) \Gamma(\eta\hat{k}_2+P,\eta\hat{k}_2) \Lambda_2(\eta\hat{k}_2)
\Biggr] \, .
\label{eq:CST-BS}
\end{align}
It determines an (approximate) BS vertex function, where both quark momenta, $p_1$ and $p_2$, are off-shell, in terms of four CST vertex functions, which always have one quark momentum on mass shell. A diagrammatic representation of Eq.~(\ref{eq:CST-BS}) is given in Fig.~\ref{fig:CST-BS}.

These CST vertex functions can be calculated, once (\ref{eq:CST-BS}) is converted into a closed set of equations.
To do so, one writes (\ref{eq:CST-BS}) for four combinations of external quark momenta, where in each case either quark 1 or 2 is on its positive or negative energy mass shell. 
We introduce the shorthand
\begin{align}
\Gamma_{1\rho}(p) & \equiv \Gamma (\rho\, \hat{p}_1,\rho\,\hat{p}_1-P) \, , \nonumber \\
 \Gamma_{2\rho}(p) & \equiv\Gamma (\rho\, \hat{p}_2 +P, \rho\, \hat{p}_2) 
\label{eq:shvf}
\end{align}
for the CST vertex functions, where $\rho = \pm$.

The corresponding four external relative momenta that appear as arguments of the kernel are 
 $p\rightarrow \{\hat{p}_1-P/2,\, \hat{p}_2+P/2,\, -\hat{p}_1-P/2,\, -\hat{p}_2+P/2 \} $,  with $\hat{p}_i=(E_{ip},{\bf p})$, and we define abbreviations for the kernel matrix elements
\begin{align}
&\mathcal{V}_{1\rho,1\eta}(p,k)\equiv \mathcal{V}(\rho\,\hat{p}_1-P/2, \eta\,\hat{k}_1-P/2),\nonumber\\
&\mathcal{V}_{1\rho,2\eta}(p,k)\equiv \mathcal{V}(\rho\,\hat{p}_1-P/2, \eta\,\hat{k}_2+P/2),
\nonumber\\
&\mathcal{V}_{2\rho,1\eta}(p,k)\equiv \mathcal{V}(\rho\,\hat{p}_2+P/2, \eta\,\hat{k}_1-P/2),\nonumber\\
&\mathcal{V}_{2\rho,2\eta}(p,k)\equiv \mathcal{V}(\rho\,\hat{p}_2+P/2, \eta\,\hat{k}_2+P/2)\, .
\label{eq:shkernel}
\end{align}
The same notation is adopted for the corresponding functions $V^K(p,k)$ that are part of the respective kernels.

\begin{figure}[tb]
\begin{center}
     \includegraphics[width=.48\textwidth]{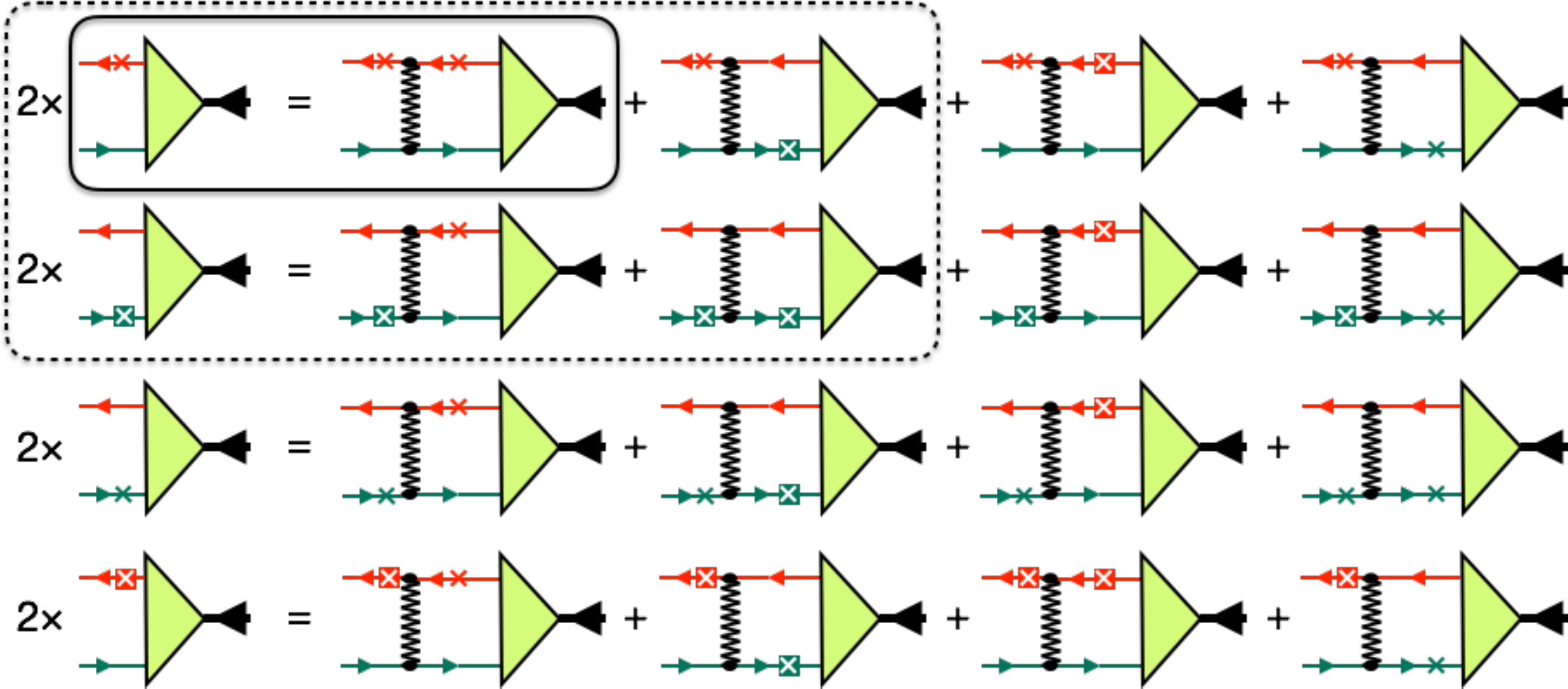}
\end{center}
  \caption{The four-channel CST equation. The solid rectangle indicates the one-channel equation used in this work, the dashed rectangle a two-channel extension with charge-conjugation symmetry.}
  \label{fig:1-2-4-CSE}
\end{figure}

Using (\ref{eq:shvf}) and (\ref{eq:shkernel}) in (\ref{eq:CST-BS}) leads to a system of four coupled equations, which we refer to as the ``four-channel CST equations" (4CSE),
\begin{align}
  \Gamma_{i \rho}&(p) = -\frac{1}{2}  \sum_{\eta=\pm} 
   \nonumber \\
   & 
\biggl[ \int_{{\bf k}_1}\mathcal{V}_{i \rho,1\eta}(p,k)\Lambda_1(\eta\hat{k}_1)\Gamma_{1\eta}(k) S_2(\eta\hat{k}_1-P )   \nonumber  \\
& + \int_{{\bf k}_2} \mathcal{V}_{i\rho,2\eta}(p,k)S_1(\eta\hat{k}_2+P ) \Gamma_{2\eta}(k)\Lambda_2(\eta\hat{k}_2)\biggr],
\label{eq:4CSTcompact}
\end{align}
where $i=1,2$, and $\rho=\pm$.
The set of equations (\ref{eq:4CSTcompact}), also shown graphically in Fig.~\ref{fig:1-2-4-CSE}, is the most general CST bound-state equation valid for quark-antiquark systems with unequal quark masses $m_1\neq m_2$, such as the heavy-light mesons that are the subject of this work.

Our interaction kernel is chosen to be of the form
\begin{multline}
{\cal V} (p,k)=\\
 \left[ (1-y) \left({\bf 1}_1\otimes {\bf 1}_2 + \gamma^5_1 \otimes \gamma^5_2 \right) - y\, \gamma^\mu_1 \otimes \gamma_{\mu 2} \right]V_\mathrm{L}(p,k) 
\\
-\gamma^\mu_1 \otimes \gamma_{\mu 2} \left[ V_\mathrm{OGE}(p,k)+V_\mathrm{C}(p,k) \right] \, ,
\label{eq:kernel}
\end{multline}
where
 $V_\mathrm{L}(p,k)$ is a covariant generalization of a linear confining potential, $V_\mathrm{OGE}(p,k)$ is the short-range one-gluon-exchange interaction (in Feynman gauge), and $V_\mathrm{C}(p,k)$ a covariant form of a constant potential. 
The OGE and constant kernels are Lorentz-vector interactions. The Lorentz structure of the linear confining kernel in (\ref{eq:kernel}) is a mixture of an equal-weight sum of scalar and pseudoscalar coupling on one hand, and vector coupling on the other hand. Our particular scalar+pseudoscalar combination ensures that the requirements of chiral symmetry are satisfied \cite{Biernat:2014uq}.
The parameter $y$ allows us to vary the relative weight of these structures continuously, with $y=0$ yielding a pure scalar+pseudoscalar coupling, and $y=1$ a pure vector coupling. 
The signs are chosen such that always---for any value of $y$---the same nonrelativistic limit is obtained, which in coordinate space corresponds to the Cornell-type potential $V(r)=\sigma r -\alpha_s/r -C$. 

For a better understanding of the nature of confinement, it is of great importance to establish the Lorentz structure of the confining interaction. In principle one can do that by treating $y$ as a free parameter that should be determined by fitting the experimental data. In Sec.~\ref{sec:results} we discuss in some detail to what extend this approach works in practice.

\subsection{Four- and two-channel equations for CST wave functions}

To bring the 4CSE (\ref{eq:4CSTcompact}) into a form more suitable for numerical solution, we begin by calculating matrix elements between $\rho$-spinors, which amounts to a separation into positive- and negative-energy channels. Our $\rho$-spinors are defined as
\begin{equation}
u^+_i({\bf p},\lambda) \equiv u_i({\bf p},\lambda)  \, , \qquad  u^-_i({\bf p},\lambda) \equiv v_i(-{\bf p},\lambda) \, ,
\end{equation}
where $u$ and $v$ are the Dirac  spinors in the convention of Bjorken and Drell, which are given explicitly in Eqs.~(\ref{eq:ui}) and (\ref{eq:vi}), and $\lambda$ is the helicity of quark $i$.

We can express the projectors and propagators in (\ref{eq:4CSTcompact}) in terms of these $\rho$-spinors as
\begin{equation}
\Lambda_i(\eta \hat{k}_i)=\eta \sum_{\lambda=\pm\frac12} u_i^\eta(\eta {\bf k}, \lambda ) \bar{u}_i^\eta(\eta {\bf k}, \lambda ),\label{eq:Lambdaexpand}
\end{equation}
and
\begin{align}
S_1(\eta \hat{k}_2 + P) &=\frac{m_1}{E_{1k}} \sum_{\rho_1=\pm }\sum_{\lambda_1=\pm\frac12} \frac{u_1^{\rho_1}(\eta{\bf k},\lambda_1) \bar{u}_1^{\rho_1}(\eta{\bf k}, \lambda_1 )}{\rho_1 E_{1k}- \eta E_{2k} - \mu-i\epsilon} \, ,\nonumber \\
S_2(\eta \hat{k}_1 - P) &=\frac{m_2}{E_{2k}} \sum_{\rho_2=\pm }\sum_{\lambda_2=\pm\frac12} \frac{u_2^{\rho_2}(\eta{\bf k},\lambda_2) \bar{u}_2^{\rho_2}(\eta{\bf k}, \lambda_2 )}{\rho_2 E_{2k}- \eta E_{1k}+ \mu-i\epsilon} \, ,
\label{eq:Sexpand}
\end{align}
respectively.

Multiplying in (\ref{eq:4CSTcompact}) $\Gamma_{1 \rho_1}$ from the left by $\bar u_1^{\rho_1}(\rho_1{\bf p}, \lambda_1)$ and from the right by $u_2^{\rho_2}(\rho_1{\bf p}, \lambda_2)$, and $\Gamma_{2 \rho_2}$ from the left by $\bar u_1^{\rho_1}(\rho_2{\bf p}, \lambda_1)$ and from the right by $u_2^{\rho_2}(\rho_2{\bf p}, \lambda_2)$, we get
\begin{widetext}
\begin{align}
 \bar u_1^{\rho_1}(\rho_1{\bf p}, \lambda_1) & \Gamma_{1 \rho_1} (p) u_2^{\rho_2}(\rho_1{\bf p}, \lambda_2) 
  = 
 -\frac 12 \sum_{K \eta \lambda'_1 \lambda'_2  \rho'_1 \rho'_2} \int \frac{\mathrm d^3 k}{(2\pi)^3}\frac{m_1 m_2}{E_{1k} E_{2k}} 
 \eta \Biggl[
 V^K_{1 \rho_1,1\eta}(p,k)
  \bar u_1^{\rho_1}(\rho_1{\bf p}, \lambda_1) {\Theta}^K_1  u_1^{\eta}(\eta{\bf k}, \lambda'_1) 
  \nonumber \\
 &   \qquad\qquad\qquad\qquad\qquad\qquad\qquad\qquad \times 
 \frac{ \bar u_1^{\eta}(\eta{\bf k}, \lambda'_1) \Gamma_{1 \eta} (k) u_2^{\rho'_2}(\eta{\bf k},\lambda'_2)}
 {\rho'_2 E_{2k}  -\eta E_{1k}+\mu -i\epsilon}
\bar u_2^{\rho'_2}(\eta{\bf k},\lambda'_2) {\Theta}^K_2  u_2^{\rho_2}(\rho_1{\bf p}, \lambda_2) 
 \nonumber \\
 & + 
 V^K_{1 \rho_1, 2\eta}(p,k)
  \bar u_1^{\rho_1}(\rho_1{\bf p}, \lambda_1) {\Theta}^K_1  u_1^{\rho'_1}(\eta{\bf k}, \lambda'_1)
   \frac{ \bar u_1^{\rho'_1}(\eta{\bf k}, \lambda'_1) \Gamma_{2 \eta} (k) u_2^{\eta}(\eta{\bf k},\lambda'_2)}
 {\rho'_1 E_{1k}  -\eta E_{2k}-\mu -i\epsilon}  
 \bar u_2^{\eta}(\eta{\bf k},\lambda'_2) {\Theta}^K_2  u_2^{\rho_2}(\rho_1{\bf p}, \lambda_2)
 \Biggr]
\nonumber \\ %%
  \bar u_1^{\rho_1}(\rho_2{\bf p}, \lambda_1) & \Gamma_{2 \rho_2} (p) u_2^{\rho_2}(\rho_2{\bf p}, \lambda_2) 
  = 
 -\frac 12 \sum_{K\eta \lambda'_1 \lambda'_2 \rho'_1 \rho'_2} \int \frac{\mathrm d^3 k}{(2\pi)^3}\frac{m_1 m_2}{E_{1k} E_{2k}} 
 \eta \Biggl[
 V^K_{2 \rho_2,1\eta}(p,k)
  \bar u_1^{\rho_1}(\rho_2{\bf p}, \lambda_1) {\Theta}^K_1  u_1^{\eta}(\eta{\bf k}, \lambda'_1) 
  \nonumber \\
 &   \qquad\qquad\qquad\qquad\qquad\qquad\qquad\qquad \times 
 \frac{ \bar u_1^{\eta}(\eta{\bf k}, \lambda'_1) \Gamma_{1 \eta} (k) u_2^{\rho'_2}(\eta{\bf k},\lambda'_2)}
 {\rho'_2 E_{2k}  -\eta E_{1k}+\mu -i\epsilon}
\bar u_2^{\rho'_2}(\eta{\bf k},\lambda'_2) {\Theta}^K_2  u_2^{\rho_2}(\rho_2{\bf p}, \lambda_2) 
 \nonumber \\
 & + 
 V^K_{2 \rho_2, 2\eta}(p,k)
  \bar u_1^{\rho_1}(\rho_2{\bf p}, \lambda_1) {\Theta}^K_1  u_1^{\rho'_1}(\eta{\bf k}, \lambda'_1)
   \frac{ \bar u_1^{\rho'_1}(\eta{\bf k}, \lambda'_1) \Gamma_{2 \eta} (k) u_2^{\eta}(\eta{\bf k},\lambda'_2)}
{\rho'_1 E_{1k}  -\eta E_{2k}-\mu -i\epsilon}  
 \bar u_2^{\eta}(\eta{\bf k},\lambda'_2) {\Theta}^K_2  u_2^{\rho_2}(\rho_2{\bf p}, \lambda_2)
 \Biggr] \, ,
   \end{align}
where the notation for the functions $ V^K_{i \rho,j\eta}(p,k)$ follows the convention of Eq.~(\ref{eq:shkernel}).
Note that repeated indices are not automatically summed over. 

Now we define CST wave functions,
\begin{align}
 \Psi_{1,\lambda_1 \lambda_2}^{\rho_1\rho_2} ({\bf p}) 
 & \equiv 
 \sqrt{\frac{m_1m_2}{E_{1p}E_{2p}}}
  \frac{ \bar u_1^{\rho_1}(\rho_1{\bf p}, \lambda_1) \Gamma_{1 \rho_1} (p) u_2^{\rho_2}(\rho_1{\bf p},\lambda_2)}
 {\rho_2 E_{2p}  -\rho_1 E_{1p}+\mu -i\epsilon}  \, ,
 \nonumber \\
  \Psi_{2,\lambda_1 \lambda_2}^{\rho_1\rho_2} ({\bf p}) 
 & \equiv 
 \sqrt{\frac{m_1m_2}{E_{1p}E_{2p}}}
  \frac{ \bar u_1^{\rho_1}(\rho_2{\bf p}, \lambda_1) \Gamma_{2 \rho_2} (p) u_2^{\rho_2}(\rho_2{\bf p},\lambda_2)}
 {\rho_1 E_{1p}  -\rho_2 E_{2p}-\mu -i\epsilon} \, ,
 \label{eq:CSTwfs}
\end{align}
and the spinor matrix elements of the vertices,
\begin{align}
\Theta^{K,\rho\rho'}_{i,\lambda\lambda'}({\bf p},{\bf k}) & \equiv 
\bar{u}_i^{\rho}({\bf p},\lambda)\Theta^K_i  u_i^{\rho'}({\bf k},\lambda')
\, .
\label{eq:thetaO}
\end{align}
The 4CSE for the CST wave functions is then
\begin{align}
 (\rho_2 E_{2p}-\rho_1 E_{1p}+\mu) \Psi_{1,\lambda_1 \lambda_2}^{\rho_1\rho_2} ({\bf p}) 
 & =
- \frac 12 \sum_{K \eta \lambda'_1 \lambda'_2  \rho'_1 \rho'_2} \int \frac{\mathrm d^3 k}{(2\pi)^3}N_{12}(p,k) \eta 
\Biggl[
 V^K_{1 \rho_1,1\eta}(p,k) \Theta^{K,\rho_1\eta}_{1,\lambda_1\lambda'_1}(\rho_1{\bf p},\eta{\bf k})
 \nonumber\\
 &\hspace{-1cm} \times 
  \Psi_{1,\lambda'_1 \lambda'_2}^{\eta\rho'_2} ({\bf k})
  \Theta^{K,\rho'_2\rho_2}_{2,\lambda'_2\lambda_2}(\eta{\bf k},\rho_1{\bf p})
  +
  V^K_{1 \rho_1,2\eta}(p,k) \Theta^{K,\rho_1\rho'_1}_{1,\lambda_1\lambda'_1}(\rho_1{\bf p},\eta{\bf k})
  \Psi_{2,\lambda'_1 \lambda'_2}^{\rho'_1\eta} ({\bf k})
  \Theta^{K,\eta\rho_2}_{2,\lambda'_2\lambda_2}(\eta{\bf k},\rho_1{\bf p}) \Biggr] \nonumber \\
   (\rho_1 E_{1p}-\rho_2 E_{2p}-\mu) \Psi_{2,\lambda_1 \lambda_2}^{\rho_1\rho_2} ({\bf p}) 
 & =
- \frac 12 \sum_{K \eta \lambda'_1 \lambda'_2  \rho'_1 \rho'_2} \int \frac{\mathrm d^3 k}{(2\pi)^3}N_{12}(p,k)
 \eta  \Biggl[
 V^K_{2 \rho_2,1\eta}(p,k) \Theta^{K,\rho_1\eta}_{1,\lambda_1\lambda'_1}(\rho_2{\bf p},\eta{\bf k})
 \nonumber\\
 & \hspace{-1cm}\times
  \Psi_{1,\lambda'_1 \lambda'_2}^{\eta\rho'_2} ({\bf k})
  \Theta^{K,\rho'_2\rho_2}_{2,\lambda'_2\lambda_2}(\eta{\bf k},\rho_2{\bf p})
  +
  V^K_{2 \rho_2,2\eta}(p,k) \Theta^{K,\rho_1\rho'_1}_{1,\lambda_1\lambda'_1}(\rho_2{\bf p},\eta{\bf k})
  \Psi_{2,\lambda'_1 \lambda'_2}^{\rho'_1\eta} ({\bf k})
  \Theta^{K,\eta\rho_2}_{2,\lambda'_2\lambda_2}(\eta{\bf k},\rho_2{\bf p}) \Biggr] \, ,
  \label{eq:4CSEwf}
\end{align}
\end{widetext}
where we have introduced the shorthand
\begin{equation}
N_{12}(p,k) \equiv \frac{m_1 m_2}{\sqrt{E_{1p} E_{2p}E_{1k} E_{2k}}} \, .
\end{equation}
To avoid potential confusion we should point out that the number of ``channels'', e.g.\ the 4 in 4CSE, refers to the number of different vertex functions $\Gamma_{i\rho}$ coupled in Eq.~(\ref{eq:4CSTcompact}), not to the total number of different $\rho$-spin components of the wave function, which is 8 in the case of Eq.~(\ref{eq:4CSEwf}).

Equation (\ref{eq:4CSEwf}) should be used when both positive-energy poles of the quark propagators contribute at a comparable level to the $k_0$ loop integration and the system is symmetric under charge conjugation. The most important example of this case is the pion.
When the total bound state mass $\mu$ is not small compared to the masses of its constituents, one pole dominates (by convention the one of particle 1), and leaving the second one out becomes a good approximation. The 4CSE (\ref{eq:4CSTcompact}) reduces then to the two-channel Covariant Spectator Equation (2CSE) 
\begin{widetext}
\begin{align}
  \Gamma_{1+ }(p) & = -\frac{1}{2}  \biggl[ 
  \int_{{\bf k}_1}\mathcal{V}_{1+,1+}(p,k)\Lambda_1(\hat{k}_1)\Gamma_{1+}(k) S_2(\hat{k}_1-P )  
  + \int_{{\bf k}_2} \mathcal{V}_{1+,2-}(p,k)S_1(-\hat{k}_2+P ) \Gamma_{2-}(k)\Lambda_2(-\hat{k}_2)\biggr]
  \nonumber \\
  \Gamma_{2- }(p) & = -\frac{1}{2}  \biggl[ 
  \int_{{\bf k}_1}\mathcal{V}_{2-,1+}(p,k)\Lambda_1(\hat{k}_1)\Gamma_{1+}(k) S_2(\hat{k}_1-P )  
  + \int_{{\bf k}_2} \mathcal{V}_{2-,2-}(p,k)S_1(-\hat{k}_2+P ) \Gamma_{2-}(k)\Lambda_2(-\hat{k}_2)\biggr] \, ,
\label{eq:2CSTcompact}
\end{align}
which couples $\Gamma_{1+ }$ with its charge-conjugation counterpart $\Gamma_{2- }$. A graphical representation of this set of equations is indicated by the dashed rectangle in Fig.~\ref{fig:1-2-4-CSE}.

The corresponding 2CSE for the CST wave function is
\begin{align}
 (\rho_2 E_{2p}- E_{1p}+\mu) \Psi_{1,\lambda_1 \lambda_2}^{+\rho_2} ({\bf p}) 
 & =
- \frac 12 \sum_{K  \lambda'_1 \lambda'_2  \rho'_1 \rho'_2} \int \frac{\mathrm d^3 k}{(2\pi)^3}N_{12}(p,k)  
\Biggl[
 V^K_{1 +,1+}(p,k) \Theta^{K,++}_{1,\lambda_1\lambda'_1}({\bf p},{\bf k})
 \nonumber\\
 &\hspace{-1.5cm} \times 
  \Psi_{1,\lambda'_1 \lambda'_2}^{+\rho'_2} ({\bf k})
  \Theta^{K,\rho'_2\rho_2}_{2,\lambda'_2\lambda_2}({\bf k},{\bf p})
  -
  V^K_{1+,2-}(p,k) \Theta^{K,+\rho'_1}_{1,\lambda_1\lambda'_1}({\bf p},-{\bf k})
  \Psi_{2,\lambda'_1 \lambda'_2}^{\rho'_1-} ({\bf k})
  \Theta^{K,-\rho_2}_{2,\lambda'_2\lambda_2}(-{\bf k},{\bf p}) \Biggr] \nonumber \\
   (\rho_1 E_{1p}+ E_{2p}-\mu) \Psi_{2,\lambda_1 \lambda_2}^{\rho_1-} ({\bf p}) 
 & =
- \frac 12 \sum_{K  \lambda'_1 \lambda'_2  \rho'_1 \rho'_2} \int \frac{\mathrm d^3 k}{(2\pi)^3}N_{12}(p,k)
  \Biggl[
 V^K_{2 -,1+}(p,k) \Theta^{K,\rho_1+}_{1,\lambda_1\lambda'_1}(-{\bf p},{\bf k})
 \nonumber\\
 & \hspace{-1.5cm}\times
  \Psi_{1,\lambda'_1 \lambda'_2}^{+\rho'_2} ({\bf k})
  \Theta^{K,\rho'_2-}_{2,\lambda'_2\lambda_2}({\bf k},-{\bf p})
  -
  V^K_{2-,2-}(p,k) \Theta^{K,\rho_1\rho'_1}_{1,\lambda_1\lambda'_1}(-{\bf p},-{\bf k})
  \Psi_{2,\lambda'_1 \lambda'_2}^{\rho'_1-} ({\bf k})
  \Theta^{K,--}_{2,\lambda'_2\lambda_2}(-{\bf k},-{\bf p}) \Biggr] \, .
  \label{eq:2CSEwf}
\end{align}

\end{widetext}

\subsection{The one-channel CST equation}

If the total bound-state mass is not small, and we are dealing with a system of particles with unequal masses, then keeping only the positive-energy pole of the heavier particle is a very good approximation. There is no need for a symmetrization as in the case of the 2CSE because charge conjugation is not a symmetry of the system. We arrive at the one-channel Covariant Spectator Equation (1CSE) for the vertex function,
 \begin{align}
  \Gamma_{1+ }(p)  = - 
  \int_{{\bf k}_1}\mathcal{V}_{1+,1+}(p,k)\Lambda_1(\hat{k}_1)\Gamma_{1+}(k) S_2(\hat{k}_1-P )  \, ,
\label{eq:1CSTcompact}
\end{align}
and the corresponding 1CSE for the CST wave function
\begin{align}
 & ( \rho_2  E_{2p}-  E_{1p}+\mu) \Psi_{1,\lambda_1 \lambda_2}^{+\rho_2} ({\bf p}) 
 =
- \hspace{-3mm} \sum_{K  \lambda'_1 \lambda'_2  \rho'_2 } \int \frac{\mathrm d^3 k}{(2\pi)^3}N_{12}(p,k) \nonumber  \\
& \hspace{6mm} \times 
 V^K_{1 +,1+}({\bf p},{\bf k}) \Theta^{K,++}_{1,\lambda_1\lambda'_1}({\bf p},{\bf k})
  \Psi_{1,\lambda'_1 \lambda'_2}^{+\rho'_2} ({\bf k})
  \Theta^{K,\rho'_2\rho_2}_{2,\lambda'_2\lambda_2}({\bf k},{\bf p}) \, . 
  \label{eq:1CSEwf}
\end{align}
The 1CSE is shown graphically inside the solid rectangle in Fig.~\ref{fig:1-2-4-CSE}. 
It is particularly well suited for heavy-light mesons, i.e.\ quark-antiquark systems with one light and one bottom or charm quark. It should also work well for heavy quarkonia, except that no definite $C$-parity can be assigned to the solutions because of the missing charge-conjugation symmetry. As we will argue in more detail in Sec.~\ref{sec:results}, in heavy quarkonia this is actually only a minor problem. It turns out that the singularity structure of the kernel matrix element in Eq.~(\ref{eq:1CSEwf}) is so much simpler than the ones that appear in the 2CSE (\ref{eq:2CSEwf}), that we consider the loss of charge-conjugation symmetry a small price to pay for the great advantages it brings with respect to its practical solution. Therefore, in this work we perform all calculations of heavy and heavy-light mesons with the 1CSE.

In the calculations of this paper, the functions $ V^K_{1 +,1+}({\bf p},{\bf k})$ that describe the momentum dependence of the various pieces of the kernel are
\begin{widetext}
\begin{multline}
V_\mathrm{L} (\hat{p}_1-P/2,  \hat{k}_1-P/2) = 
-8\sigma \pi\Biggl[\left(\frac{1}{(\hat{p}_1-\hat{k}_1)^4}-\frac{1}{\Lambda^4+(\hat{p}_1-\hat{k}_1)^4}\right) 
\\
  -\frac{E_{1p}}{m_1}(2\pi)^3 \delta^3 (\mathbf{p-k})
\int_{{\bf k}_1'}\left(\frac{1}{(\hat{p}_1-\hat{k}_1')^4}-\frac{1}{\Lambda^4+(\hat{p}_1-\hat{k}_1')^4}\right)\Biggr]\,,
\label{eq:VL}
\end{multline}
for the linear confining kernel, assumed equal for scalar ($K=s$), pseudoscalar ($K=p$), and vector coupling ($K=v$),
\begin{equation}
V_\mathrm{OGE}(\hat{p}_1-P/2,  \hat{k}_1-P/2)  = -4 \pi \alpha_s \left(\frac{1}{(\hat{p}_1-\hat{k}_1)^2}-\frac{1}{(\hat{p}_1-\hat{k}_1)^2-\Lambda^2}\right) \,, 
\label{eq:VOGE}
\end{equation} 
for the one-gluon exchange (in Feynman gauge), and
\begin{equation}
V_\mathrm{C}(\hat{p}_1-P/2,  \hat{k}_1-P/2) = (2\pi)^3\frac{E_{1k}}{m_1} C \delta^3 (\mathbf{p-k}) \, .
\label{eq:VC}
\end{equation}
\end{widetext}
for the covariant generalization of a constant kernel, the latter two both in vector coupling ($K=v$).
The three constants $\sigma$, $\alpha_s$, and $C$ are the adjustable coupling strength parameters of the interaction model. The confining and OGE kernels in (\ref{eq:VL}) and  (\ref{eq:VOGE}) are shown in Pauli-Villars regularized form, which introduces the cutoff parameter $\Lambda$. Without regularization, the loop integration in (\ref{eq:1CSEwf}) would not converge.

To solve Eq.~(\ref{eq:1CSEwf}) numerically, we represent the wave functions in a basis of eigenfunctions of the 
total orbital angular momentum $L$ and total spin $S$ of the quark-antiquark system. Although neither $L$ nor $S$ are conserved quantum numbers, this is useful when we want to compare our results to nonrelativistic approaches which classify their states in terms of $L$ and $S$. It is also interesting to get a  measure of the importance of relativistic effects by quantifying the extent to which partial waves of purely relativistic origin mix with the ones present in nonrelativistic theories.

For this purpose, the wave functions (\ref{eq:CSTwfs}) and kernel matrix elements (\ref{eq:thetaO}) in (\ref{eq:4CSEwf}) are written as matrix elements of the two-component helicity spinors $\chi_\lambda$, using the spinor representation defined in Eqs.~(\ref{eq:ui}) and (\ref{eq:vi}). In the remainder of this section  $p$ and $k$ refer to the magnitudes of the three-vectors $\bf p$ and $\bf k$, and should not  be mistaken as  four-vectors. 

We write the kernel vertex matrix elements as 
\begin{align}
\Theta^{K,\rho\rho'}_{i,\lambda\lambda'}({\bf p},{\bf k}) =
N_{ip} N_{ik} \chi^\dagger_\lambda M_i^{K,\rho\rho'}({\bf p},{\bf k}) \chi_{\lambda'}
\, ,
\label{eq:thetaO2}
\end{align}
where $N_{ip}=\sqrt{\frac{E_{ip}+m_i}{2m_i}}$, and the $2\times 2$ matrices $M^K_i$ depend on the Lorentz structure of the vertex specified by the superscript $K$. All matrix elements needed for the Lorentz structure of the kernel (\ref{eq:kernel}) are listed in Appendix~\ref{app:B}.

Similarly, the wave functions are written as
\begin{align}
& \Psi^{+\rho}_{1,\lambda\lambda'} ({\bf p})=\sum_{j} \psi_j^\rho(p) \chi^\dagger_\lambda\, K_j^\rho(\hat{\bf p}) \, \chi_{\lambda'} \, ,
\label{eq:Psidesomp}
\end{align}
where $\hat{\bf p}$ is a unit vector in the direction of ${\bf p}$, and the index $j$ distinguishes linearly independent matrices $K_j^\rho(\hat{\bf p})$, which we choose such that each term in the sum (\ref{eq:Psidesomp}) corresponds to a quark-antiquark 
eigenstate of $L$ and $S$. The matrix representation (\ref{eq:Psidesomp}) is  interpreted as describing quark 2 entering the vertex and quark 1 coming out of it, as shown in Fig.~\ref{fig:BSE}, whereas eigenstates of $L$ and $S$ refer to linear combinations of direct product states describing a quark and an antiquark both leaving the vertex. The latter involve sums over Clebsch-Gordan coefficients and spherical harmonics, which will then appear in the matrices $K_j^\rho(\hat{\bf p})$ when the direct product representation is transformed into the matrix representation. An example of the relation between the two representations can be found in Ref.~\cite{Buck:1979vn}. 

Equation (\ref{eq:Psidesomp}) represents therefore a partial wave decomposition of the CST wave function, where $\psi_j^\rho(p)$ are radial wave functions, and the spin and angular dependence is contained in the matrices $K_j^\rho(\hat{\bf p})$. For $J^P=0^-$ mesons, there is only one independent matrix for each value of $\rho$, namely an $S$-wave for $\rho=-$, and a $P$-wave for $\rho=+$. The $1^-$ mesons have two different matrices $K_j^\rho$ for each value of $\rho$, namely an $S$ and a $D$ wave for $\rho=-$, and spin singlet and triplet $P$ waves for $\rho=+$. For $0^+$ and $1^+$ mesons, the respective partial waves in $\rho=+$ and $\rho=-$ are interchanged. The explicit expressions of $K_j^\rho(\hat{\bf p})$ are given in Appendix~\ref{app:A}.  

After inserting the expansion (\ref{eq:Psidesomp}) into Eq.~(\ref{eq:1CSEwf}), and using the completeness of the $\chi_\lambda$-spinors, the bound state equation takes on the form
\begin{align}
& (\rho E_{2p}-E_{1p}+\mu)\sum_{j} \psi_{j}^{\rho}(p)  K_{j}^{\rho}(\hat{\bf p}) 
= -\int \frac{d^3k}{(2\pi)^3} N(p,k)
\nonumber \\
& 
\times  \sum_{K \rho' j'} V^K({\bf p},{\bf k}) 
  M_1^{K,++}({\bf p},{\bf k})
 \, \psi_{j'}^{\rho'}(k) K_{j'}^{\rho'}(\hat{\bf k}) 
 M_{2}^{K,\rho'\rho}({\bf k},{\bf p}) \, ,
 \label{eq:1CSEa}
\end{align}
with $ N(p,k) \equiv N_{1p} N_{1k}N_{2k} N_{2p} N_{12}(p,k) $.

We can simplify Eq.~(\ref{eq:1CSEa}) by using the fact that the kernel $V^K$ depends only on the magnitudes of the three-vectors $\bf p$ and $\bf k$ and on the angle between them, i.e.,
\begin{equation}
V^K({\bf p},{\bf k}) = V^K(p,k,z) \, ,
\end{equation}
 where $p=|{\bf p}|$, $k=|{\bf k}|$, and $z=\hat{\bf p} \cdot \hat{\bf k}$. In general, if $f(p,k,z)$ is a function of this kind, one can determine new functions $A_{j j'}^{K,\rho\rho'}(p,k,z)$ such that
 \begin{align}
&\int \frac{d^3k}{(2\pi)^3}  f(p,k,z)  M_1^{K,++}({\bf p},{\bf k}) K_{j'}^{\rho'}(\hat{\bf k}) M_{2}^{K,\rho'\rho}({\bf k},{\bf p}) 
\nonumber \\
 &=
\int \frac{d^3k}{(2\pi)^3}  f(p,k,z)  \sum_{ j} K_{j}^{\rho}(\hat{\bf p}) A_{j j'}^{K,\rho\rho'}(p,k,z)    \, .
\end{align}
Using this relation in  (\ref{eq:1CSEa}), we obtain

\begin{align}
&  (\rho E_{2p}-E_{1p}+\mu)\sum_{j} \psi_{j}^{\rho}(p)  K_{j}^{\rho}(\hat{\bf p}) 
=  - \hspace{-2mm} \sum_{K  j \rho' j'} \int \frac{d^3k}{(2\pi)^3}
\nonumber \\
 &  \times N(p,k) 
 V^K(p,k,z)  K_{j}^{\rho}(\hat{\bf p}) A_{j j'}^{K,\rho\rho'}(p,k,z) \psi_{j'}^{\rho'} (k) \, .
 \label{eq:1CSEc}
\end{align}
Matrices $K_{j'}^{\rho'}(\hat{\bf p})$ belonging to different orbital angular momenta are orthogonal with respect to integration over $\hat{\bf p}$, whereas spin singlet and triplet matrices are orthogonal with respect to taking the trace of their product. One can therefore extract an equation for the coefficients of these matrices in (\ref{eq:1CSEc}), which can be written
\begin{align}
&  (E_{1p}-\rho E_{2p})\psi_{j}^{\rho}(p)
  - \sum_{K \rho' j'} \int \frac{d^3k}{(2\pi)^3} N(p,k) 
 V^K(p,k,z)
\nonumber \\
 & \hspace{20mm} \times    A_{j j'}^{K,\rho\rho'}(p,k,z) \psi_{j'}^{\rho'} (k) =
 \mu \, \psi_{j}^{\rho}(p) \, .
 \label{eq:1CSEd}
\end{align}
This is a linear eigenvalue equation whose eigenvalues $\mu$ are the bound state masses, and the corresponding eigenvectors are the radial partial wave functions $\psi_{j}^{\rho}(p)$. 

It is one of the great advantages of the 1CSE that the integrand in (\ref{eq:1CSEd}) itself does not  depend explicitly on $\mu$. Solving this equation yields the ground state and a tower of excited states at once. A dependence of the integrand on $\mu$ usually turns the equation into a nonlinear problem, where one has to search for a self-consistent solution for each eigenvalue separately. In the 1CSE this is the case, for instance, when the fixed constituent quark mass of the off-shell quark is replaced by a dynamical mass function, and it is unavoidable in the 2CSE and 4CSE even for fixed quark masses.

We have solved Eq.~(\ref{eq:1CSEd})  by expanding the wave functions $\psi_{j}^{\rho}(p)$ in a basis of B-splines. The numerical methods, and in particular the way how a linear confining interaction and its covariant generalization can be treated in momentum space, have been described in some detail in Refs.~\cite{GMilana:1994,Uzzo,Leitao:2014}. 

Once the partial wave functions $\psi_j^\rho(p)$ have been calculated, we can also construct the vertex functions $\Gamma(p_1,p_2)$.  If $\Gamma(p_1,p_2)$ is written in terms of covariant Lorentz tensors multiplied by functions of invariants $G_{l} (p_1^2,p_2^2)$ as  in Appendix~\ref{sec:covstruc}, Eqs.~(\ref{eq:Psidesomp}) and (\ref{eq:CSTwfs}) relate the $\psi$'s with the $G$'s. In many applications, the vertex function in this manifestly covariant form is more useful.

\section{Numerical results}
\label{sec:results}

In a recent letter \cite{Leitao:2017it} we presented first results of our calculations of the masses of heavy and heavy-light mesons with $J=0$ and $1$, based on the 1CSE (\ref{eq:1CSEd}). We performed least square fits of the three kernel parameters $\sigma$, $\alpha_s$, and $C$, while choosing fixed values for the constituent quark masses and an equal-weight scalar and pseudoscalar coupling for the confining interaction (i.e., with $y=0$). The Pauli-Villars cutoff parameter was fixed at $\Lambda= 2 m_1$ (we also used this choice in the new results presented below). We found that the obtained models describe the experimental masses very well, with an rms difference between calculations and data of the order of 30 MeV.

In this work we extend our previous study in several aspects:

(i) The parameter $y$ describing the mixing of scalar/pseudoscalar and vector confining interaction is promoted to an adjustable parameter. One of the most interesting questions we want to investigate is of course whether the meson mass spectrum can determine $y$ or at least yield useful constraints.

(ii) We also treat now the constituent quark masses as adjustable parameters. This may seem a rather straightforward way to improve the fits of \cite{Leitao:2017it}. However, it represents a serious complication in the required numerical calculations: the interaction kernel depends linearly on the constants $\sigma$, $\alpha_s$,  $C$, and $y$, and the most time-consuming part of the calculation, namely the loop-momentum integration in (\ref{eq:1CSEd}) needs to be carried out only once. On the other hand, the kernel's dependence on the quark masses is much more complicated. When the quark masses are allowed to vary, this numerical integration over the kernel has to be recalculated every time the combination of masses is changed during the fits.  

(iii) In \cite{Leitao:2017it} we found that a fit of the coupling constants exclusively to pseudoscalar meson masses gives overall results that are almost as good as when additionally vector and scalar states are also used in the fit. Here we explore how much our results depend on the selection of the fitted data set in the new, more general fits.

(iv) Although they are not observables, it is useful to have a closer look at the relativistic ``wave functions''. They provide a means to identify the quantum numbers of the corresponding bound states. In the case of heavy mesons, we expect the dominant component to closely resemble a corresponding nonrelativistic wave function. 
The weight of the wave function components of relativistic origin should increase with decreasing quark mass. Their sensitivity to changes in the model parameters will be explored as well.

\subsection{Interaction models and mass spectra}
\label{sec:spectra}
We calculated the pseudoscalar, scalar, vector, and axialvector meson states that contain at least one heavy (bottom or charm) quark, and whose mass falls below the corresponding open-flavor threshold. As exceptions, a few states located
slightly above threshold but with very small widths are considered as well. We restrict our analysis to mesons with $J^P=0^\pm, 1^\pm$, representing already the vast majority of the experimental states.

\begin{table}[tb]{
\centering
\begin{tabular}{lll|lccc}
 & & & \multicolumn{3}{c}{Data set} \\
State        & $J^{P(C)}$ & Mass (MeV)     & S1 & S2 & S3  \\
\hline
$\Upsilon (4S)$        & $1^{--}$         & 10579.4$\pm$1.2    &      & $\bullet$ &   $\bullet$ \\
$\chi_{b1} (3P)$      & $1^{++}$         & 10512.1$\pm$2.3     &     & &   $\bullet$ \\
$\Upsilon (3S)$        & $1^{--}$         & 10355.2$\pm$0.5    &      & $\bullet$ &  $\bullet$   \\
$\eta_b(3S)$           & $0^{-+}$         & 10337        &  &  \\
$h_{b} (2P)$            & $1^{+-}$         & 10259.8$\pm$1.2                          &     & &   $\bullet$  \\
$\chi_{b1} (2P)$      & $1^{++}$         & 10255.46$\pm$0.22$\pm$0.50     &     & &   $\bullet$  \\
$\chi_{b0} (2P)$      & $0^{++}$         & 10232.5$\pm$0.4$\pm$0.5     &     & $\bullet$ &  $\bullet$  &  \\
$\Upsilon (1D)$         & $1^{--}$         &  10155  &   &  \\
$\Upsilon (2S)$        & $1^{--}$         & 10023.26$\pm$0.31    &    & $\bullet$ &  $\bullet$  \\
$\eta_b(2S)$           & $0^{-+}$         & 9999$\pm$4                               & $\bullet$ &   $\bullet$ &  $\bullet$  \\
$h_{b} (1P)$            & $1^{+-}$         & 9899.3$\pm$0.8                          &    & &   $\bullet$  \\
$\chi_{b1} (1P)$      & $1^{++}$         & 9892.78$\pm$0.26$\pm$0.31     &   & &   $\bullet$ \\
$\chi_{b0} (1P)$      & $0^{++}$         & 9859.44$\pm$0.42$\pm$0.31     &    & $\bullet$ &  $\bullet$  \\
$\Upsilon (1S)$        & $1^{--}$         & 9460.30$\pm$0.26      &   & $\bullet$ &  $\bullet$ \\
$\eta_b(1S)$           & $0^{-+}$         & 9399.0$\pm$2.3          & $\bullet$ &   $\bullet$ &  $\bullet$   \\
\hline
$B_c(2S)^\pm$           & $0^{-}$         & 6842$\pm$6          &    & &   $\bullet$  \\
$B^+_c$           & $0^{-}$         & 6275.1$\pm$1.0          & $\bullet$ &   $\bullet$ &  $\bullet$  \\
\hline
$B_{s1}(5830)$           & $1^{+}$         & 5828.63$\pm$0.27         &     & &   $\bullet$  \\
$B_1(5721)^{+,0}$         & $1^{+}$         & 5725.85$\pm$1.3       &    & &   $\bullet$  \\
$B^*_s$           & $1^{-}$         & 5415.8$\pm$1.5         &     & $\bullet$ &  $\bullet$   \\
$B^0_s$           & $0^{-}$         & 5366.82$\pm$0.22          & $\bullet$ &   $\bullet$ &  $\bullet$  \\
$B^*$           & $1^{-}$         & 5324.65$\pm$0.25         &      & $\bullet$ &  $\bullet$   \\
$B^{\pm,0}$          & $0^{-}$          & 5279.45      & $\bullet$ &   $\bullet$ &  $\bullet$  \\
\hline
$X(3915)$     & $0^{++}$         & 3918.4$\pm$1.9     &     & $\bullet$ &  $\bullet$   \\
$\psi (3770)$        & $1^{--}$         & 3773.13$\pm$0.35    &      & $\bullet$ &  $\bullet$   \\
$\psi (2S)$        & $1^{--}$         & 3686.097$\pm$0.010    &      & $\bullet$ &  $\bullet$   \\
$\eta_c(2S)$           & $0^{-+}$         & 3639.2$\pm$1.2                               & $\bullet$ &   $\bullet$ &  $\bullet$  \\
$h_{c} (1P)$            & $1^{+-}$         & 3525.38$\pm$0.11                          &     & &   $\bullet$  \\
$\chi_{c1} (1P)$      & $1^{++}$         & 3510.66$\pm$0.07     &    & &   $\bullet$  \\
$\chi_{c0} (1P)$      & $0^{++}$         & 3414.75$\pm$0.31     &     & $\bullet$ &  $\bullet$   \\
$J/\Psi (1S)$        & $1^{--}$         & 3096.900$\pm$0.006      &     & $\bullet$ &  $\bullet$   \\
$\eta_c(1S)$           & $0^{-+}$         & 2983.4$\pm$0.5         & $\bullet$ &   $\bullet$ &  $\bullet$  \\
\hline
$D_{s1}(2536)^{\pm}$           & $1^{+}$         & 2535.10$\pm$0.06         &     & &   $\bullet$  \\
$D_{s1}(2460)^{\pm}$           & $1^{+}$         & 2459.5$\pm$0.6         &     & &   $\bullet$  \\
$D_1(2420)^{\pm,0}$         & $1^{+}$         & 2421.4       &     & &   $\bullet$  \\
$D^*_{0}(2400)^{0}$           & $0^{+}$         & 2318$\pm$29         &      & $\bullet$ &  $\bullet$  \\
$D^{*}_{s0}(2317)^\pm$           & $0^{+}$         & 2317.7$\pm$0.6         &     & $\bullet$ &  $\bullet$   \\
$D^{*\pm}_s$           & $1^{-}$         & 2112.1$\pm$0.4          &     & $\bullet$ &  $\bullet$   \\
$D^*(2007)^0$         & $1^{-}$         & 2008.62       &     & &   $\bullet$  \\
$D^\pm_s$           & $0^{-}$         & 1968.27$\pm$0.10          & $\bullet$ &   $\bullet$ &  $\bullet$  \\
$D^{\pm,0}$          & $0^{-}$          & 1867.23      & $\bullet$ &   $\bullet$ &  $\bullet$  \\
\end{tabular}%
}
\caption{List of the mesonic states and experimental measured masses used throughout this work. A bullet point in one of the columns labeled S1, S2, and S3 indicates that the meson state is included in the respective data set used in various fits. The masses of $B^{\pm,0}$, $D^{\pm,0}$, $B_1(5721)^{+,0}$, and  $D_1(2420)^{\pm,0}$ are averages of the charged and uncharged states. The masses of $\Upsilon(1D)$ and $\eta_b(3S)$ are estimates taken from Ref.~\cite{Godfrey:2014fk}. There is weak evidence (at $1.8\sigma$) that $\Upsilon(1D)$ has been seen \cite{Bonvicini:2004fr,Amo-Sanchez:2010jk}.}
\label{tab:expMass}
\end{table}

There are two different ways how we quantify the relation between the masses $\mu_i(\{\alpha_k(M)\})$, calculated from a theoretical model $M$ specified through a set of parameters $\{\alpha_k(M)\}$, and a certain set $S$ of  experimental masses $\mu_i^\mathrm{exp}(S)$ with $N_S$ elements. When $S$ is the set of data used in the least square fit of the model parameters, then the rms difference
\begin{equation}
\delta_\mathrm{rms}(S)\equiv
 \sqrt{\frac{1}{N_S}\sum_{i\in S} \left[\mu_i(\{\alpha_k(M)\}) -\mu_i^\mathrm{exp}(S)\right]^2 }
\end{equation}
is the quantity that is being minimized, and its value is therefore a measure of the quality of the fit.

\begin{table*}[htb]
\begin{center}
\begin{tabular}{  c c | c c c c cccc| c c c  }
  Model &symbol & $\sigma$ (GeV$^2$) & $\alpha_s$ &$C$ (GeV) & $y$ & $m_b$ (GeV) & $m_c$ (GeV) &$m_s$ (GeV) &$m_q$ (GeV) & $N$ & $\delta_\mathrm{rms}$ (GeV) & $\Delta_\mathrm{rms}$ (GeV)\\ \hline
  M0$_{\text{S1}}$   &    & 0.2493 & 0.3643 & 0.3491 & {\bf 0.0000} & {\bf 4.892}& {\bf 1.600} &{\bf 0.4478} & {\bf 0.3455} & 9 & 0.017& 0.037\\
    M1$_{\text{S1}}$&  \color{blue} $\bigcirc$
&0.2235 & 0.3941 & 0.0591  &0.0000 & 4.768 & 1.398& 0.2547& 0.1230& 9  & 0.006 &0.041 \\
    \hline
    M0$_{\text{S2}}$& & 0.2247 & 0.3614 & 0.3377  & {\bf 0.0000} & {\bf 4.892} & {\bf 1.600} & {\bf 0.4478}& {\bf 0.3455} & 25  &0.028 &0.036 \\
M1$_{\text{S2}}$& & 0.1893 & 0.4126 & 0.1085  &0.2537 & 4.825 & 1.470& 0.2349& 0.1000 & 25  &0.022 & 0.033\\
\hline
M1$_{\text{S2}'}$& \color{red} $\triangle$& 0.2017 & 0.4013 & 0.1311 & 0.2677 &  4.822 & 1.464 & 0.2365 & 0.1000 &  24& 0.018 & 0.033 \\
\hline
M1$_{\text{S3}}$&${\fcolorbox{Green}{white!!white}{}}$  & 0.2022 & 0.4129 & 0.2145  &0.2002 & 4.875 & 1.553 & 0.3679 & 0.2493 & 39  &0.030 & 0.030 \\
M0$_{\text{S3}}$& ${\fcolorbox{black}{green!!white}{}}$  & 0.2058 & 0.4172 & 0.2821  &{\bf 0.0000}  & 4.917 & 1.624 & 0.4616 & 0.3514 & 39& 0.031 & 0.031
    \end{tabular}
\end{center}
    \caption{Summary table of the kernel parameters of the different fitting models considered in this work. The masses calculated from the models labeled with the symbols  {\color{blue}$\bigcirc$} , {\color{red}$\triangle$}, $^{\fcolorbox{Green}{white!!white}{}}$  and $^{\fcolorbox{black}{green!!white}{}}$ are shown in Fig.\,\ref{fig:spectrum}. $N$ is the number of states in the data set used in fitting the model. $\delta_\mathrm{rms}$ indicates the minimized root mean square difference with respect to the data set used in the fit, and $\Delta_\mathrm{rms}$ is the root mean square difference with respect to data set S3, including both fitted and predicted states. The values in boldface were held fixed. }
    \label{tab:parameters}
    \end{table*}
\begin{figure*}[bht]
%\begin{center}
\includegraphics[width=0.90\textwidth]{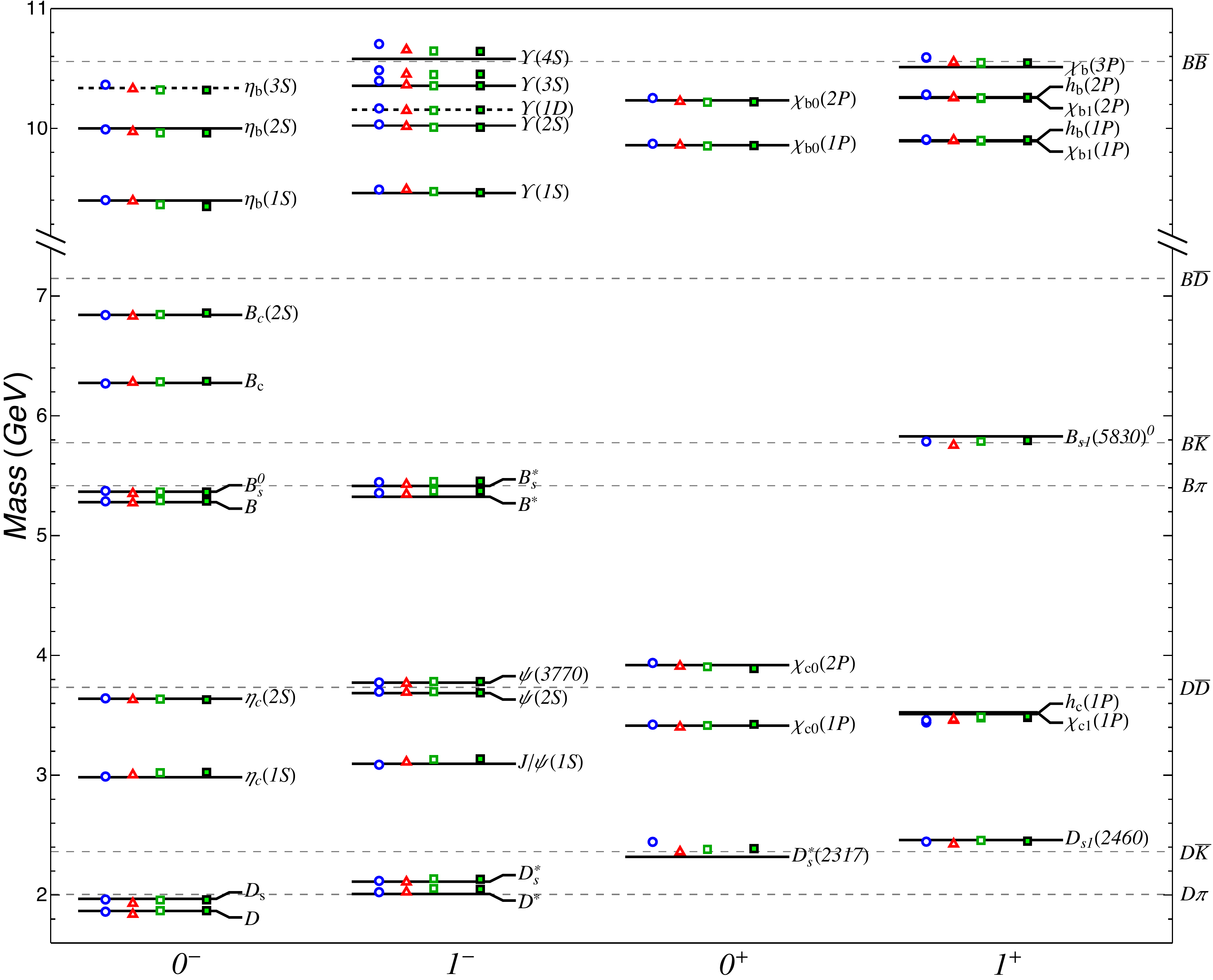}
\caption{Masses of heavy-light and heavy mesons with $J^P=0^\pm$ and $1^\pm$. The points depicted by the symbols {\color{blue}$\bigcirc$} , {\color{red}$\triangle$}, ${\fcolorbox{Green}{white!!white}{}}$  and $^{\fcolorbox{black}{green!!white}{}}$ represent the 1CSE results calculated with the models with matching symbols of Table II. Solid horizontal lines are the measured meson masses \cite{PDG2014}. The two dashed levels are estimates taken from Ref.~\cite{Godfrey:2014fk}. There is weak evidence (at $1.8\sigma$) that the $\Upsilon(1D)$ has been seen \cite{Bonvicini:2004fr,Amo-Sanchez:2010jk}. All models predict a so far unobserved $\Upsilon(2D)$ between $\Upsilon(3S)$ and $\Upsilon(4S)$.  Dashed horizontal lines across the figure indicate open flavor thresholds.}
\label{fig:spectrum}
%\end{center}
\end{figure*}

On the other hand, we also want to be able to evaluate the ability of a given model to predict states it was not fitted to. 
For this purpose we also calculate rms differences with respect to data sets $S'$ that are different from the set $S$ a model was fitted to. To distinguish these differences more clearly from the minimized values we use the notation $\Delta_\mathrm{rms}(S')$ whenever $S' \ne S$.
Note that it is quite possible that, for particular choices of $S$ and $S'$, one model has a higher $\delta_\mathrm{rms}$ but a smaller $\Delta_\mathrm{rms}$ than another.

We chose three different sets of data to fit our model parameters to:  the set called S1 consists of pseudoscalar meson states only (it is identical to the one used in \cite{Leitao:2017it} to fit the model named P1), the set S2 includes pseudoscalar, scalar, and vector states, and the largest set, S3, adds a number of axial vector states to the states contained in S2. A  list of these states and their masses is given in Table~\ref{tab:expMass}.

We constructed several interaction models by fitting to these three data sets while, in some cases, placing constraints on certain parameters. The results of our fits are summarized in Table~\ref{tab:parameters}. In all cases, the rms difference $\Delta_\mathrm{rms}$ is given with respect to the data set S3, containing a total of 39 states.

Models M0$_{\text{S1}}$ and M0$_{\text{S2}}$, previously denoted in ref.\cite{Leitao:2017it} by P1 and PSV1 respectively, were fitted with fixed values for the constituent quark masses and mixing parameter ${y=0}$ \cite{Leitao:2017it}. They should be compared to the new models M1$_{\text{S1}}$ and M1$_{\text{S2}}$, in which the quark masses and $y$ were allowed to vary freely. We see that the addition of 5 free parameters leads to a lower minimum in $\delta_\mathrm{rms}$, but the overall rms difference $\Delta_\mathrm{rms}$ changes by very little (it even increases from M0$_{\text{S1}}$ to M1$_{\text{S1}}$). Based on the data set S1, the fit finds no improvement in varying $y$, such that the new minimum is located again at $y=0$. This is not the case for data set S2, which prefers a finite value of $y$ of approximately $0.25$. At the same time, the quark masses change quite considerably, decreasing by around 200 MeV (more moderately for $m_b$), which is in part compensated by a similarly smaller constant $C$. To see that this compensating effect makes sense, remember that $q\overline{q}$ spinor matrix elements of $\gamma^\mu_1 \otimes \gamma_{\mu 2}$ are negative in the dominant channel with $\rho'=-$. Because of the overall minus sign in the definition of $V_C(p,k)$, lowering $C$ makes the kernel on the rhs of Eq.~(\ref{eq:1CSEc}) smaller, and lowering the quark masses reduces its lhs. 
The masses of the light quarks tend to go as low as possible in these fits. The final value of 100 MeV is actually the lower limit of the range in which they were allowed to vary.

The bottomonium system is very rich in measured excited states. This poses a bit of a challenge for our calculations, because describing higher excited states accurately requires a larger number of spline functions. In particular, the $\Upsilon$(4S) appears in our calculations as the 5th excited state in the vector $b\overline{b}$ system, but increasing the number of basis spline functions accordingly would be too time-consuming to perform our 8-parameter fits. To test whether the M1$_{\text{S2}}$ fit might have been distorted by trying to reproduce the $\Upsilon$(4S) mass with insufficient numerical accuracy, we performed another fit where this state was omitted from the fitted data set. To distinguish from the previous one we denote it by S2$'$. However, the resulting model, M1$_{\text{S2}'}$, turned out very similar to M1$_{\text{S2}}$, and produces the same value of $\Delta_\mathrm{rms}$. 

Finally, we fitted two more models to our largest data set, S3, which adds axial vector mesons to the set S2. The parameters $\sigma$ and $\alpha_s$ of M1$_{\text{S3}}$ are quite similar to those of M1$_{\text{S2}}$, but the quark masses are all higher, which is again accompanied by an increase of the constant $C$. The mixing parameter turns out a bit smaller, at $y=0.20$. To see how sensitive the fit is to the precise value of $y$, we repeat the calculation with the same data set, but with the restriction $y=0$. The coupling strength parameters of this model, M0$_{\text{S3}}$, are almost unchanged compared to M1$_{\text{S3}}$, only the quark masses (and $C$) increase.
It is reassuring that, in both cases, the light quark masses have moved back into a more realistic region, around 300 MeV.

The overall quality of these fits is slightly better than the one of all previous models. We consider M1$_{\text{S3}}$, with $\Delta_\mathrm{rms}=0.030$ GeV, our best model. But the fact that for M0$_{\text{S3}}$ the rms difference $\Delta_\mathrm{rms}=0.031$ GeV is only marginally larger is a strong indication that the parameter $y$ is not significantly constrained by the heavy and heavy-light meson spectrum, at least not by the states in the data sets we used. We will study this point in more detail in the next section.

Figure~\ref{fig:spectrum} compares the meson masses calculated with models M1$_{\text{S1}}$, M1$_{\text{S2}^\prime}$, M1$_{\text{S3}}$, and M0$_{\text{S3}}$, with the experimental data \cite{PDG2014}. The overall agreement is very good in all cases. It is remarkable that model M1$_{\text{S1}}$, whose parameters were determined by fits to pseudoscalar states only, yields results of almost the same quality as the other models. As we discussed in  \cite{Leitao:2017it}, this implies that requiring the kernel to be of covariant form correctly determines the spin-dependent interactions, which are responsible for the splitting between the different $J^{P(C)}$ channels. 
It is worth emphasizing that ours are global fits, where the same parameters are used in all sectors of the shown spectrum. This is in contrast to other models frequently found in the literature that adjust their parameters sector by sector in order to achieve a better fit.

As already discussed, the 1CSE is ideally suited for the description of heavy and heavy-light mesons, i.e.\ when at least  one constituent is a charm or bottom quark. However, one drawback of the 1CSE is that it is not symmetric under charge conjugation. Consequently we cannot assign a definite $C$-parity to our solutions for heavy quarkonia. 

This issue becomes relevant only in the case of axial vector mesons, which come in both $C$-parities. The observed splitting between these $C$-parity pairs is very small, about 5 MeV in bottomonium and 14 MeV in charmonium, and the $C=+$ state is always the one lower in mass. Our solutions of the 1CSE yield also closely spaced pairs in the $J^P=1^+$ channel. The problem is that, when performing a fit, we need to know which calculated state should be compared to which experimental one. It is quite possible that, when regions in the parameter space far from the final minimum are probed, the ordering of states in the calculated spectrum is not equal to the experimental one, which could lead to incorrect identifications and potentially drag the fit away from the true minimum.

In practice there are mitigating circumstances that essentially eliminate this problem. The first is that heavy quarkonia are close to the nonrelativistic limit, especially bottomonium. 
Relativistically, both spin singlet ($S=0$) and spin triplet ($S=1$) configurations may contribute to a state of definite $C$ parity and orbital angular momentum $L$. This is different from the nonrelativistic limit, where the relation $C=(-)^{L+S}$ holds, implying that either one or the other of the two spin states goes to zero for a given $C$ parity. For instance, if $L=1$, $S=0$ does not contribute to the $C=+$ state, and  $S=1$ does not contribute to the  $C=-$ state in the nonrelativistic limit.

The CST equations have a smooth nonrelativistic limit, therefore the axialvector quarkonium wave functions should be dominated by  $P$-wave components with either $S=1$ or $S=0$, while $S$ and $D$ waves should be very small. This is indeed what we find, such that by determining whether a state has a dominant spin triplet or singlet wave function we can decide which experimental state it should be compared to.

The second aspect is that the mass splitting between the $C=+$ and $C=-$ pairs is very small indeed, actually even smaller than the numerical accuracy we estimate our numerical solutions to have. This means that even if calculated and experimental states were occasionally paired incorrectly, it would hardly have a significant influence on the fits.

\subsection{CST wave functions}

In this section we present the CST wave functions for a selection of mesons. They will be used in the future to calculate electroweak form factors and decay rates, as well as hadronic decay properties. They are also fundamental ingredients in calculations of many other hadronic reactions that involve the formation of these mesons. It is therefore of great importance to understand their structure in detail.
All wave functions displayed here were calculated with model M1$_{\text{S3}}$, and are normalized according to (\ref{eq:pnorm}), (\ref{eq:snorm}), (\ref{eq:vnorm}), and (\ref{eq:wfnorm}). 

\begin{figure*}
\centering
\includegraphics[width=14cm]{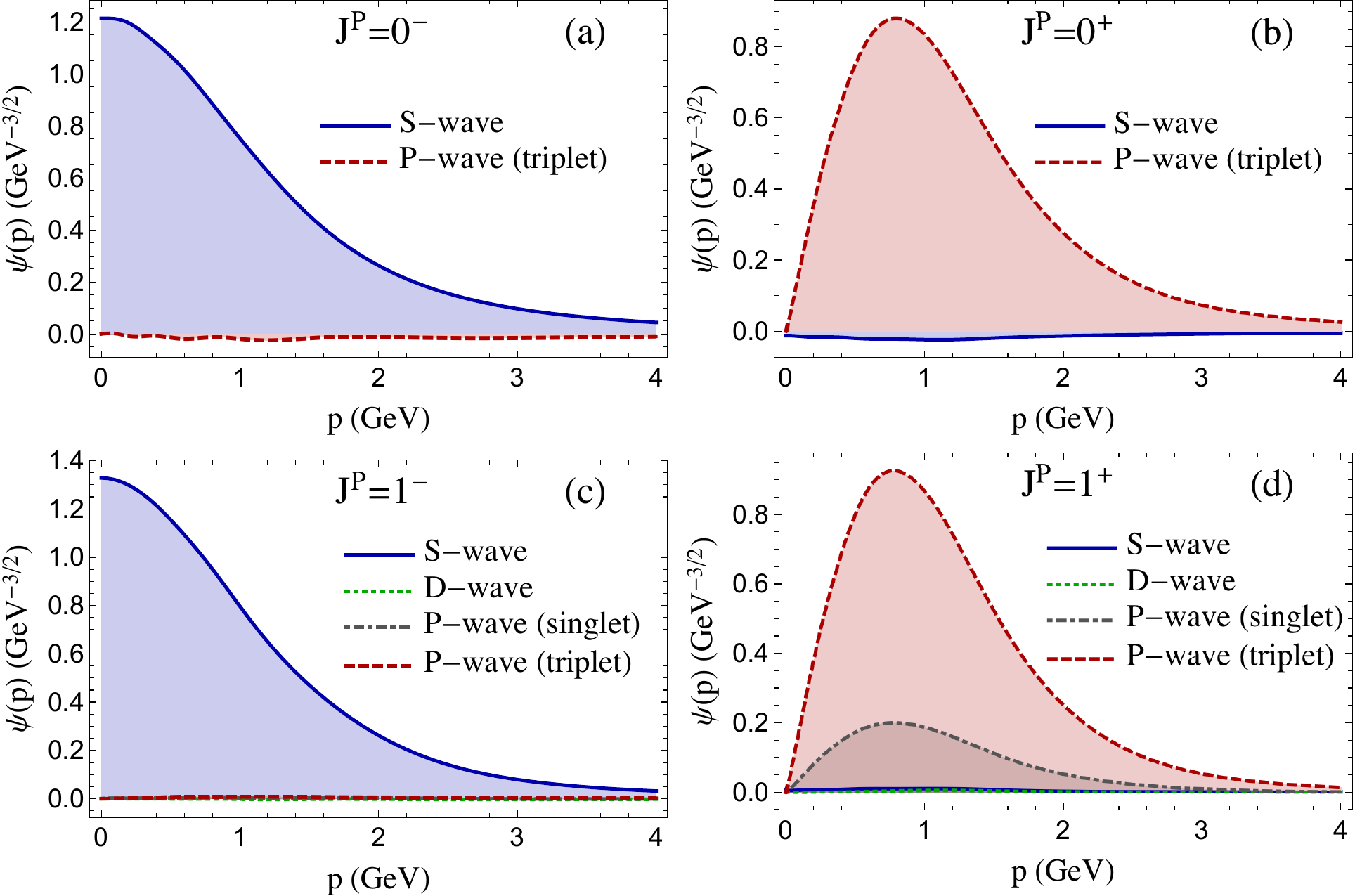}
\caption{Wave function components calculated with the parameters of model M1$_{\text{S3}}$, represented by symbol $^{{\tiny \fcolorbox{Green}{white!!white}{} } }$ in Fig.~\ref{fig:spectrum}, for the $b\bar{b}$ ground states. The corresponding meson names and quantum numbers $J^P$ are (a) $\eta_b(1S)$ with $J^P=0^{-}$,  (b) $\chi_{b0}(1P)$ with $J^P=0^{+}$,  (c) $\Upsilon(1S)$ with $J^P=1^{-}$, and (d) $\chi_{b1}(1P)$ with $J^P=1^{+}$. Solid lines represent $S$-waves, dashed lines triplet $P$-waves, dashed-dotted lines singlet $P$-waves, and dotted lines are $D$-waves.}
\label{fig:wfBB}
\end{figure*}

Figure \ref{fig:wfBB} shows the ground-state wave functions of bottomonium in the four channels $J^P=0^\pm$, $1^\pm$. The pseudoscalar and vector mesons are almost pure $S$ waves, and the scalar and axial vector mesons are almost pure $P$ waves. The weight of the components of relativistic origin is so small that their wave functions are difficult to distinguish from zero in the plots. Because of the large mass of the $b$ quark the bottomonium behaves essentially nonrelativistically.

\begin{figure*}
\centering
\includegraphics[width=14cm]{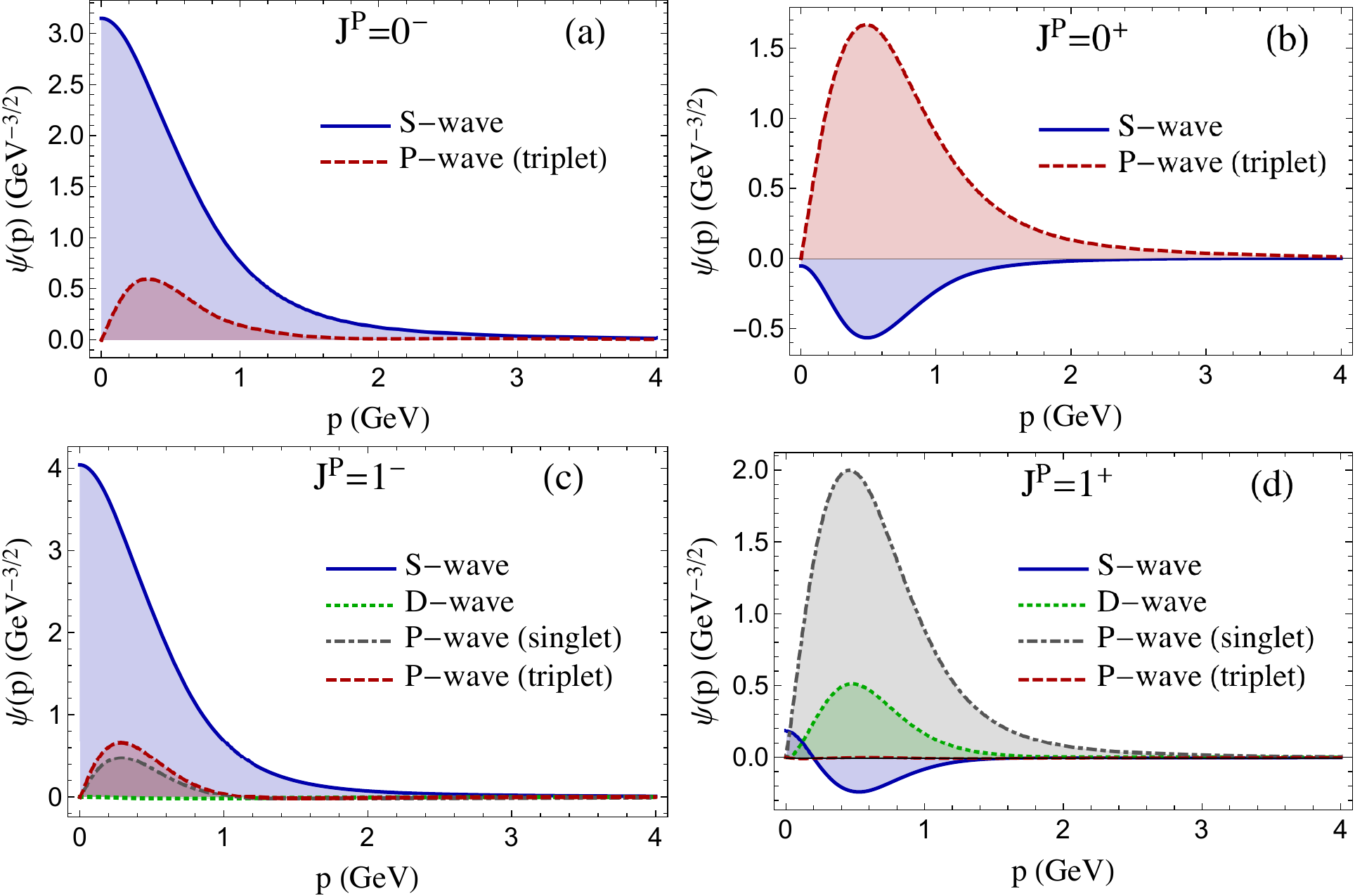}
\caption{Wave function components calculated with the parameters of model M1$_{\text{S3}}$, represented by symbol $^{{\tiny \fcolorbox{Green}{white!!white}{} } }$ in Fig.~\ref{fig:spectrum}, for the $c\bar{q}$ ground states. The corresponding meson names, where available, and quantum numbers $J^P$ are (a) $D$ with $0^{-}$,  (b) $0^{+}$,  (c) $D^*$ with $1^{-}$, and (d) $1^{+}$. The $c\bar{q}$ states (b) and (d) have not yet been observed experimentally, but are predicted by our model M1$_{\text{S3}}$ at 2.293 GeV ($0^+$) and 2.367 GeV ($1^+$). The lines have the same meaning as in Fig.\,\ref{fig:wfBB}.
}
\label{fig:wfCQ}
\end{figure*}

\begin{figure*}[tb]
\centering
\includegraphics[width=14cm]{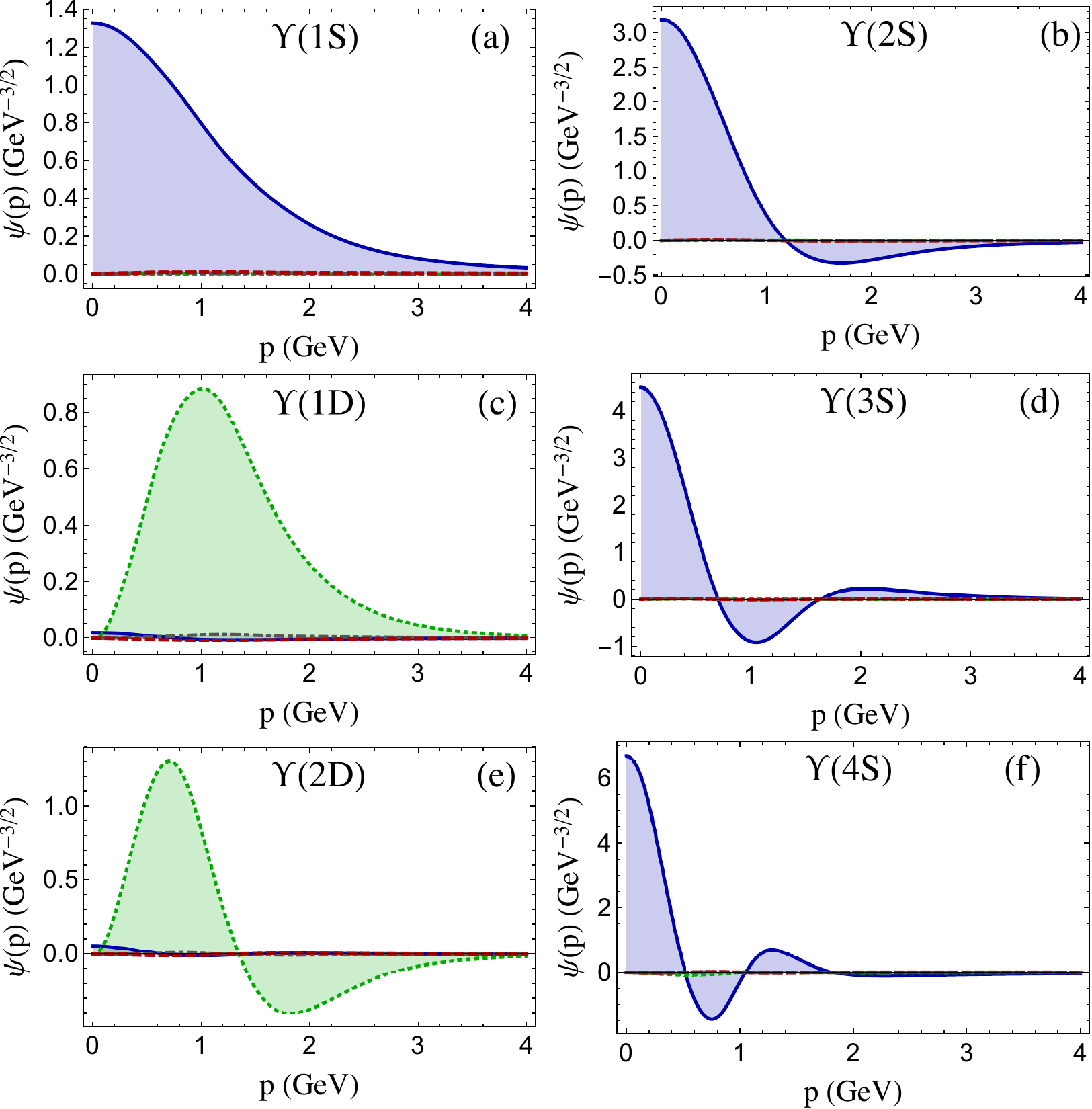}
\caption{Wave function components for the six lowest-mass states of $b\bar{b}$ with $J^{P}=1^{-}$, calculated with the parameters of model M1$_{\text{S3}}$, represented by symbol $^{{\tiny \fcolorbox{Green}{white!!white}{} } }$ in Fig.~\ref{fig:spectrum}. The order of the panels from (a) to (f) corresponds to increasing mass of the state. The lines have the same meaning as in Fig.\,\ref{fig:wfBB}.}
\label{fig:excitedBB}
\end{figure*}

One can then expect that the relativistic components are more pronounced in systems with lighter quarks. Figure \ref{fig:wfCQ} shows the  wave functions analogous to the ones in Fig.~\ref{fig:wfBB} for the lightest $c\overline{q}$ mesons ($q$ stands collectively for a light $u$ or $d$ quark, with $m_u=m_d=m_q$). As expected, the relativistic components are already quite significant, and a nonrelativistic description is no longer adequate. 

Comparing Figs.~\ref{fig:wfBB} and \ref{fig:wfCQ} one can also see that the momentum-space wave functions of bottomonium are much more spread out, which means that in configuration space they are more compact than the heavy-light $c\overline{q}$ mesons.

Figure \ref{fig:wfBB}(d) contains another interesting detail: the $1^+$ ground state is dominated not by one, but by a mixture of two $P$ waves, a spin triplet and a spin singlet. The role of these two $P$ waves is interchanged in the first excited state (not shown in the figure). As already discussed in the previous section, in a relativistic description both spin triplets and singlets can contribute to either $C$-parity eigenstate. However, the plot in Fig.~\ref{fig:wfBB}(d) may give an exaggerated impression of the weight of the singlet $P$-wave: its contribution to the total norm is actually only about 7 \%. Nevertheless, the fact that in the almost nonrelativistic $\chi_\mathrm{b1}(1P)$ the singlet component is not smaller is probably in part due to the lack of charge conjugation symmetry of the 1CSE.  We can speculate that this singlet wave function will be more suppressed when a charge-conjugation symmetric two- or four-channel CST equation is solved. In addition, the presence of a pseudoscalar confining kernel also enhances its weight. When it is turned off, the norm integral of the singlet $P$-wave is reduced by roughly one half.

The vector meson spectrum of bottomonium is particularly interesting because of the large number of excited states below or slightly above threshold that have been measured.  In Fig.~\ref{fig:excitedBB} we show the wave functions of the first six vector states of bottomonium. According to the figure, the first two states are mostly $S$ waves, followed by alternating $D$ and $S$ states. The $\Upsilon(1D)$ is listed in \cite{PDG2014} as a $2^{++}$ state, but there is some evidence that $1^{--}$ was also possibly seen. There is, however,  no experimental evidence yet for the predicted $\Upsilon(2D)$. The figure shows that there is a small mixture of $2S$ in our $\Upsilon(1D)$, and a small $3S$ component is present in the $\Upsilon(2D)$. Apart from the increasing number of nodes, one can also clearly see that the wave functions are the more concentrated at lower momenta the higher excited a state is, which means that they are increasingly spread out in configuration space. 

Whereas the structure of the ground state is determined mostly by the OGE interaction, the higher excited states should be more sensitive to the confining interaction. We have already seen in the previous section that the masses of these states can be well described by our models. To test the importance of the confining interaction for the description of the bottomonium excitation spectrum, we performed fits using the OGE and constant kernels only. The quality of these fits turned out significantly worse, with rms differences above 100 MeV, compared to about 30 MeV when the complete kernel is used. Moreover, the sequence of $S$- and $D$-wave dominated states is altered in the bottomonium vector meson spectrum: the $\Upsilon(2D)$ and $\Upsilon(4S)$ swap places. This finding suggests that, once the $\Upsilon(2D)$ is observed, finding its mass below or above the mass of $\Upsilon(4S)$ can tell us whether a linear confining interaction is indeed needed or not.

\subsection{Constraints on fit parameters}

Our model fits of Tab.~\ref{tab:parameters} show some variation in the values of the best-fit parameters, depending on which data set the model is fitted to. In this section we want to investigate this sensitivity in more detail and determine how well some of the parameters are actually constrained.

\begin{figure}[tb]
\centering
\includegraphics[width=8.5cm]{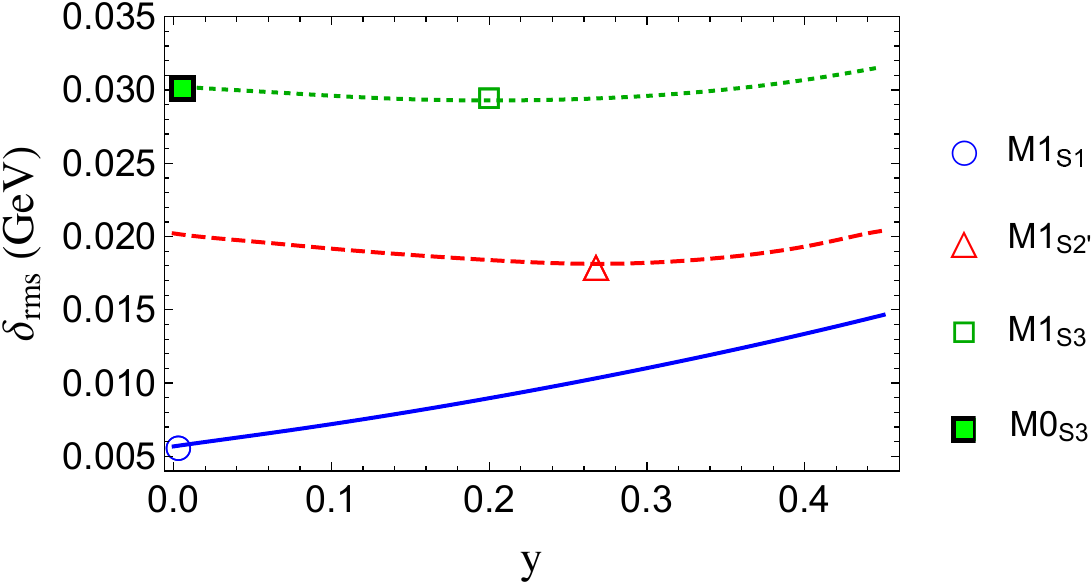}
\caption{Variation of $\delta_{\text{rms}}$ in a series of fits where the parameter $y$ has been held fixed while all other parameters were fitted. The solid line shows the result of fits to data set S1 of Tab.~\ref{tab:expMass}, the dashed and dotted lines refer to data sets S2 and S3, respectively. 
The symbols {\color{blue} $^\bigcirc$}, {\color{red} $^\triangle$}, $^{\fcolorbox{Green}{white!!white}{}}$,  and $^{\fcolorbox{black}{green!!white}{}}$ indicate the results of models M1$_{\text{S1}}$, M1$_{\text{S2}'}$, M1$_{\text{S3}}$, and M0$_{\text{S3}}$ of Tab.~\ref{tab:parameters}.}
\label{fig:crms}
\end{figure}

 \begin{figure*}[htb]
 \centering
\includegraphics[width=12.5cm]{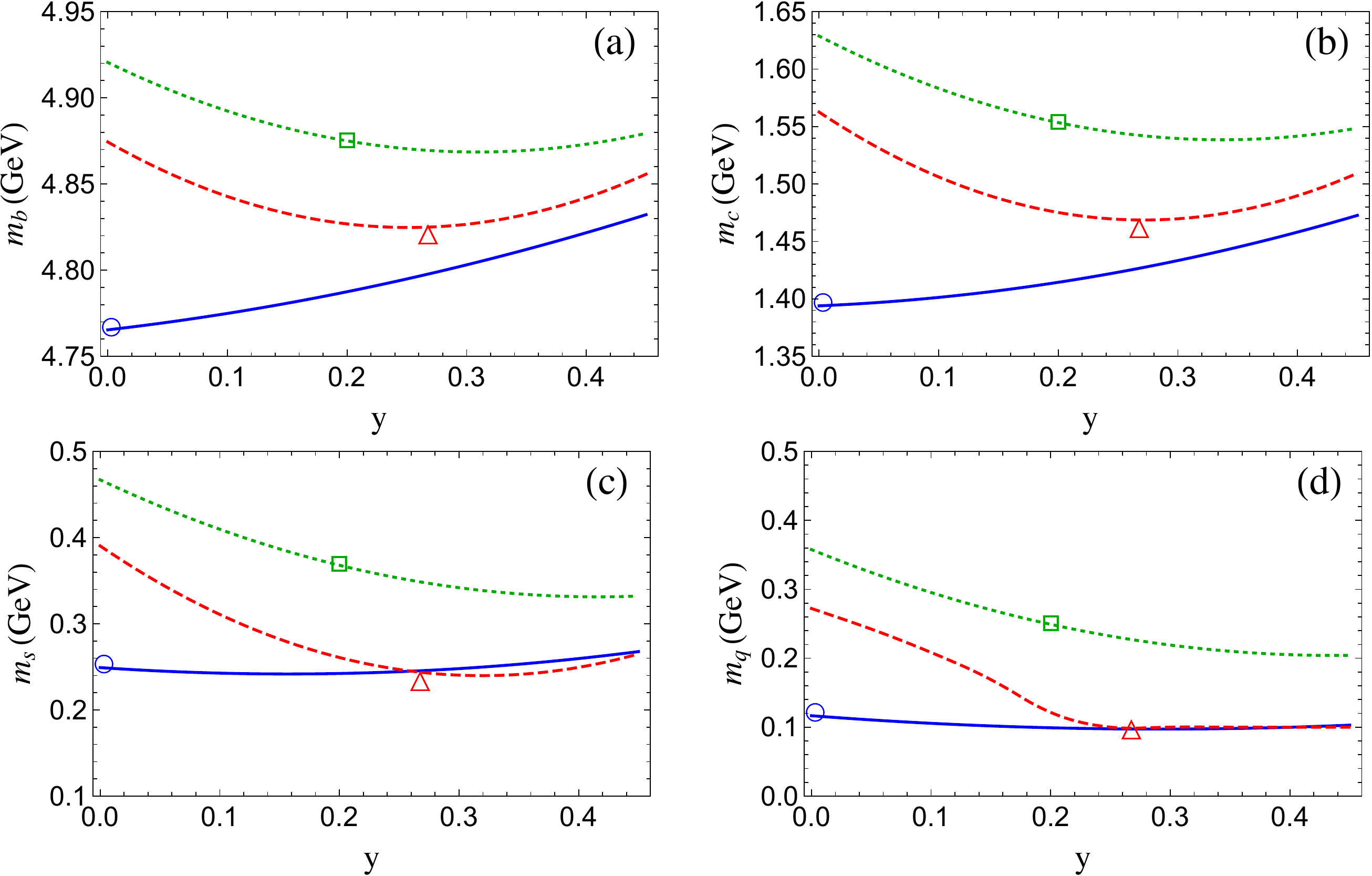}\quad
\caption{Variation of the fitted constituent quark masses in a series of fits where the parameter $y$ has been held fixed while all other parameters were fitted. The panels show the masses of (a) the bottom, (b) the charm, (c) the strange, and (d) the light (up and down) quarks, respectively. The meaning of the lines and symbols is the same as in Fig.~\ref{fig:crms}.}
\label{fig:ymasses}
\end{figure*}

 \begin{figure*}[htb]
\centering
\includegraphics[width=17.7cm]{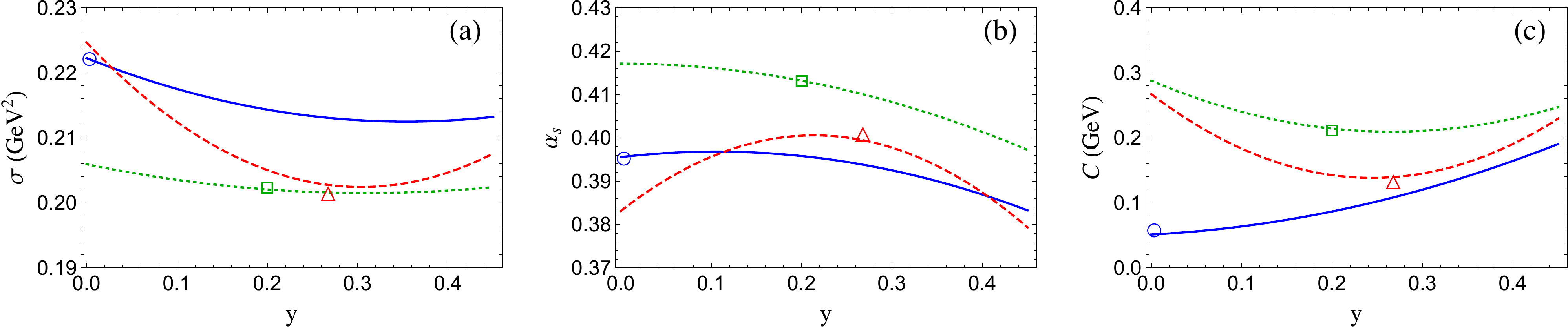}
\caption{Variation of the fitted interaction strength parameters in a series of fits where the parameter $y$ has been held fixed while all other parameters were fitted. The panels show (a) the strength of the linear confining interaction $\sigma$, (b) the OGE coupling strength $\alpha_s$, and (c) the constant $C$. The meaning of the lines and symbols is the same as in Fig.~\ref{fig:crms}.}
\label{fig:cpar}
\end{figure*}

We begin with the parameter $y$ that determines the mixing between the scalar+pseudoscalar and vector confining interaction. We perform a series of fits, where in each case $y$ is held fixed at a different value while all other parameters are allowed to vary. We restrict $y$ to lie in the interval between $0$ and $0.45$. For higher values, the equation becomes unstable and no physical solutions can be found---a well-known phenomenon that was observed with many different relativistic equations \cite{Bander1984,Uzzo}.

Figure~\ref{fig:crms} shows the obtained minima of $\delta_\mathrm{rms}$ as a function of $y$, using three
different data sets. As already discussed in Sec.~\ref{sec:spectra}, the data set with exclusively pseudoscalar mesons prefers $y=0$, whereas optimum values of $y$ between $0.20$ and $0.27$ are obtained when more data are included. However, Fig.~\ref{fig:crms} also shows that, except for the smallest data set, the minima are very shallow. In fact, when using data set S3, no particular value of $y$ seems to be clearly favored over any other. Instead of accepting the value $y=0.20$ of the fit M1$_{\text{S3}}$, we could choose arbitrarily another value without deteriorating the fit significantly. 

Figure~\ref{fig:ymasses} shows how the constituent quark masses adjust when $y$ is changed, and Fig.~\ref{fig:cpar} displays the corresponding variations of the couplings strengths parameters $\sigma$, $\alpha_s$, and $C$. For the larger data sets, a trend is visible that connects smaller $y$ with somewhat higher masses, whereas the variations in the coupling strength parameters are rather mild. Overall, the heavy quark masses stay within a range of the size of about 50 MeV, while the lighter quark masses vary by around 100 MeV. But the midpoint of that range depends also on the data set of the model fit.

We can summarize that the fits to the heavy and heavy-light mass spectra alone do not lead to a clear conclusion whether the confining interaction is of pure Lorentz scalar+pseudoscalar nature or if it includes a Lorentz vector component as well.

 \section{Summary and conclusions}
 \label{sec:conclusions}
In this work, we apply the Covariant Spectator Theory (CST) to describe mesons as relativistic quark-antiquark bound states.
We briefly review how the most general CST equations, the four-channel spectator equation (4CSE) can be derived from the Bethe-Salpeter equation, and how the two- and one-channel approximations (2CSE and 1CSE) are obtained and motivated. These are momentum-space integral equations, formulated in Minkowski space, that can be cast into the form of eigenvalue problems where the eigenvalues yield the bound-state mass spectrum and the eigenvectors are the corresponding relativistic wave functions. Our numerical method to solve these equations uses a partial-wave expansion.
We provide explicit expressions that relate our partial wave solutions to a manifestly covariant representation of the corresponding meson vertex functions. This is very practical when the vertex functions are used in the calculation of elastic or transition meson form factors, decay properties, or other reactions involving mesons. 

Heavy and heavy-light mesons are bound states in which one constituent is either a charm or bottom quark, whereas the second can be either light or heavy. The 1CSE is ideally suited to describe these systems, and it is also simple enough to let us use least-square fits to determine the optimal parameters of our models. 

We have applied the 1CSE to construct models of the quark-antiquark interaction with a kernel containing a covariant generalization of a linear confining potential, a one-gluon exchange (OGE) and a ``covariantized'' constant interaction. The confining kernel has a mixed Lorentz structure, namely an equal-weight scalar and pseudoscalar part on one hand, and a vector part on the other. The particular combination of scalar and pseudoscalar interactions satisfies the requirements of chiral symmetry \cite{Biernat:2014uq}. Its weight relative to the Lorentz vector interaction is controlled by an adjustable mixing parameter, $y$. The OGE and constant kernels are pure vector interactions.

In previous work \cite{Leitao:2017it}, we have fitted only the three coupling strength parameters to the spectrum of heavy and heavy-light mesons with $J^P=0^\pm$ and $1^\pm$, while the constituent quark masses were held fixed and the mixing parameter was set to $y=0$, corresponding to a scalar+pseudoscalar Lorentz structure without vector contribution. Here we extend this work by letting $y$ and all quark masses be determined by the fit, the latter representing a significant complication of the numerical calculations. 

We find several models that reproduce the mass spectrum of heavy and heavy-light mesons with very good accuracy, as measured by the rms difference between calculated and experimental masses. It is important to emphasize that we perform \emph{global} fits, i.e., our model parameters are the same for all mesons, not varied sector by sector. 

When we fit to pseudoscalar states only, $y=0$ is obtained as the best value, and all other meson masses are remarkably well predicted.
But when the fit is based on a more extended data set that includes pseudoscalar, scalar, and vector mesons (and axial vector mesons in the most complete cases), a 20\%-25\% contribution of vector coupling is preferred. However, we found that the minima of the corresponding rms differences as functions of $y$ are very shallow, such that a model with $y=0$ is not significantly worse than one with the best fit value. The same can be said about the dependence of the models on the quark masses. When the light quark masses ($m_u=m_d$ and $m_s$) are varied within an interval of about 100 MeV, and the heavy quark masses ($m_c$ and $m_b$) within an interval of about 50 MeV, no particular values yield clearly better fits than others. 

Our main conclusion from these calculations is that the Lorentz structure of the confining interaction cannot be determined very well through the heavy and heavy-light meson mass spectrum alone, because the mixing parameter $y$ is not sufficiently constrained by these data. Nevertheless, other physical observables of these mesons are likely to be more sensitive to $y$, for instance the decay constants, which probe details of their wave functions \cite{Leitao:2017esb}. 
Similar considerations apply to the constituent quark masses, where we find that relatively large variations are compatible with the experimental spectrum.  

We also show radial wave functions for a selection of meson states. Examining wave functions is useful to identify the quantum numbers of calculated states. Relativistic wave functions contain also partial wave components which are forbidden in a nonrelativistic framework. The norm integral of these components of purely relativistic origin can be interpreted as a measure of the importance of relativity in the description of a quark-antiquark system. As expected, we find that the weight of these partial waves is very small in heavy quarkonia, and increases when quark masses become smaller, reaching about 9\% in the case of the $c\bar q$ system.

For higher excited states the momentum-space wave functions concentrate at smaller momenta, which reflects spatially more extended systems. The accurate description of highly excited states requires considerable care with the applied numerical methods. The fact that not only the meson mass spectrum is well reproduced, but  also 
the shapes of our wave functions for the excited states look reasonable and change as one would expect, is a good indication that our numerical methods to solve the 1CSE are working reliably.

The work reported in this paper completes successfully the first stage of our larger project of constructing a self-consistent unifying framework for all mesons with a quark-antiquark structure. Already at this stage, using the one-channel CST equation, we obtain a remarkably good  description of both heavy and heavy-light sectors simultaneously. The obtained wave functions can now be used as ingredients in the calculation of a wide variety of hadronic processes and experimentally observable quantities, for instance bottomonium, charmonium, and heavy-light meson decay constants, charmonium electroweak elastic and transition form factors, such as $J/\psi \rightarrow \eta_c\, \gamma^*$, $J/\psi \rightarrow \chi_{c0} \,\gamma$, $\chi_{c1} \rightarrow J/\psi \,\gamma$ and $h_c \rightarrow \eta_{c} \,\gamma$.

\begin{acknowledgments}
We thank Franz Gross for fruitful discussions and valuable suggestions, and the Jefferson Lab Theory Group for its hospitality. 
This work was supported by Funda\c c\~ao para a Ci\^encia e a Tecnologia (FCT) under contracts SFRH/BD/92637/2013, SFRH/BPD/\-100578/\-2014, and UID/FIS/0777/2013.
\end{acknowledgments}

\begin{appendix}

\section{Covariant and partial-wave tensor bases}\label{app:A}
In this appendix, we present, for each type of meson $M=p$ (pseudoscalar), $s$ (scalar), $v$ (vector), and $a$ (axial-vector), the relations between the Lorentz-invariant functions $G_n^M(p_1^2,p_2^2)$ in the covariant expansion of the meson vertex function and the radial wave functions $\psi_j^\rho (p)$ of the partial-wave components.
%%%%%%%%%%%%%%%%%%%%%%%%%%%%%%%%%%%%%%%%%%%%%%%%%
%%%%%%%%%%%%%%%%%%%%%%%%%%%%%%%%%%%%%%%%%%%%%%%%%%
\subsection{Covariant basis}
\label{sec:covstruc}
 
%%%%%%%%%%%%%%%%%%%%%%%%%%%%%%%%%%%%%%%%%%%%%%%%%%%%%%%%%%%%%%%%%
\subsubsection{Spin-0 mesons}
The invariant vertex function $\Gamma^{M}(p_1,p_2)$ connecting two off-shell quarks with momenta $p_1$ and $p_2$ can be written for pseudoscalar and scalar mesons as \begin{eqnarray}
\Gamma^{p}(p_1,p_2)=G^p_{1}\gamma^5+G^{p}_{2}\gamma^5\Lambda_2 + \Lambda_1 G^{p}_{3}\gamma^5 
 +\Lambda_1  G^{p}_{4}\gamma^5\Lambda_2 \label{eq:tensorGammaP}\nonumber\\
\end{eqnarray} and
\begin{eqnarray}
&\Gamma^{s}(p_1,p_2)=G^s_{1}+G^{s}_{2}\Lambda_2 + \Lambda_1 G^{s}_{3}+
&\Lambda_1  G^{s}_{4}\Lambda_2 \label{eq:tensorGammaS}\,,
\end{eqnarray}
respectively. Here we have introduced the shorthand (for the Lorentz-invariant functions) $\Lambda_i\equiv\Lambda_i (-p_i)=(m_i-\slashed p_i)/2m_i $ and $G_{n}^M\equiv G_{n}^M(p_1^2,p_2^2)$.
%%%%%%%%%%%%%%%%%%%%%%%%%%%%%%%%%%%%%%%%%%%%%%%%%%%%%%%%%%%%%%%%%
\subsubsection{Spin-1 mesons}
For vector and axialvector mesons, the covariant vertex functions $\tilde\Gamma^{M\mu}(p_1,p_2)$ can be written in the general form 
\begin{eqnarray}
& \tilde \Gamma^{v\mu }&(p_1,p_2)=G^v_{1}\gamma^\mu+G^v_{5}p^\mu 
+G^v_{9}P^\mu\nonumber
 \\&&+\left(G^v_{2}\gamma^\mu+G^v_{6}p^\mu+G^v_{10}P^\mu\right)\Lambda_2\nonumber
 \\&&
 + \Lambda_1\left(G^v_{3}\gamma^\mu+G^v_{7}p^\mu+G^v_{11}P^\mu\right)\nonumber
 \\&&
+\Lambda_1\left(G^v_{4}\gamma^\mu+G^v_{8}p^\mu+G^v_{12}P^\mu\right)\Lambda_2
 \label{eq:tensorGammaV}
\end{eqnarray}
and
\begin{eqnarray}
& \tilde \Gamma^{a\mu }&(p_1,p_2)=G^a_{1}\gamma^\mu\gamma^5+G^a_{5}p^\mu\gamma^5+G^a_{9}P^\mu\gamma^5\nonumber\\&&
 +\left(G^a_{2}\gamma^\mu\gamma^5+G^a_{6}p^\mu\gamma^5+G^a_{10}P^\mu\gamma^5\right)\Lambda_2\nonumber\\&&+ \Lambda_1\left(G^a_{3}\gamma^\mu\gamma^5+G^a_{7}p^\mu\gamma^5+G^a_{11}P^\mu\gamma^5\right)
 \nonumber\\&&
 +\Lambda_1 \left(G^a_{4}\gamma^\mu\gamma^5+G^a_{8}p^\mu\gamma^5+G^a_{12}P^\mu\gamma^5\right)
 \Lambda_2\label{eq:tensorGammaA}\,,
\end{eqnarray}
respectively.
Massive spin-1 particles are transverse, satisfying $P^\mu \xi_\mu (\lambda,P)=0$, where $\xi^\mu\equiv \xi^\mu (\lambda,P)$ are the spin-1 polarization four-vectors with $\lambda=0,\pm1$. Contracting $\tilde \Gamma^{v\mu }(p_1,p_2)$ and $\tilde \Gamma^{a\mu }(p_1,p_2)$ with  $\xi_\mu$ removes the longitudinal components proportional to $P^\mu$, defining the (transverse) invariant vertex functions $\Gamma^{v}(p_1,p_2)\equiv\xi_\mu\tilde \Gamma^{v\mu}(p_1,p_2)$ and $\Gamma^{a}(p_1,p_2)\equiv\xi_\mu\tilde \Gamma^{a\mu}(p_1,p_2)$ for vector and axialvector mesons, respectively. 
\vspace*{1cm}
%%%%%%%%%%%%%%%%%%%%%%%%%%%%%%%%%%%%%%%%%%%%%%%%%%%%%%%%%%%%%%%%%
\subsection{CST wave functions and the partial-wave tensor basis}
 
We use the standard representation for the Dirac matrices and four-spinors $u_i^{\rho}$ (in the convention of Bjorken-Drell) given by 
 \begin{eqnarray}
 u_i^{+}({\bf p},\lambda) & \equiv & u_i({\bf p},\lambda)=N_{ip}
 \left(\begin{array}{c}
   {\mathbf 1}\\
   \frac{\boldsymbol\sigma\cdot {\bf p}}{E_{i p}+m_i}
 \end{array}\right)\otimes\chi_\lambda   , \label{eq:ui} \\
 u_i^{-}({\bf p},\lambda) & \equiv & v_i(-{\bf p},\lambda)=N_{ip}
 \left(\begin{array}{c}
   -\frac{\boldsymbol\sigma\cdot {\bf p}}{E_{i p}+m_i}\\ {\mathbf 1}
 \end{array}\right)\otimes\chi_\lambda , \label{eq:vi}
\end{eqnarray}
where $i=1 $ or $2$ denotes the outgoing or incoming quark, respectively, $\chi_\lambda$ are the two-component spinors, and $N_{ip}=\sqrt{\frac{E_{i p}+m_i}{2m_i}}$.

For the CST vertex functions, we introduce the shorthand 
\begin{eqnarray}
 G^{M\rho_1}_{n,1}\equiv G_{n}^M \left(m_1^2,(\rho_1 \hat p_1-P)^2\right) \, , \\
 G^{M\rho_2}_{n,2}\equiv G_{n}^M\left ( (\rho_2 \hat p_2+P)^2,m_2^2\right) \, ,
 \end{eqnarray}
 where quark 1 or quark 2 is on mass shell, respectively, with positive ($\rho_i=+$) or negative ($\rho_i=-$) energy. 

Inserting the expansions~(\ref{eq:tensorGammaP})--(\ref{eq:tensorGammaA}) for each type of meson into Eq.~(\ref{eq:CSTwfs}) gives the results listed in the subsections below. In the calculation of the corresponding spinor matrix elements of the vertex functions, we use the relations
 \begin{eqnarray} \bar u_1^{\rho_1}(\rho_2{\bf p},\lambda_1)\left[m_1-(\rho_2\hat{\slashed p}_2+\gamma^0\mu)\right]&=&\nonumber\\   &&\hspace{-4cm}\bar u_1^{\rho_1}(\rho_2{\bf p},\lambda_1)\gamma^0
 (-\mu+\rho_1 E_{1 p}-\rho_2 E_{2 p})\,,\nonumber\\   
           \left[ m_2-(\rho_1\hat{\slashed p}_1-\gamma^0\mu)\right] u_2^{\rho_2}(\rho_1{\bf p},\lambda_2)&=&\nonumber\\&&\hspace{-4cm} (\mu-\rho_1 E_{1 p}+\rho_2 E_{2 p})\gamma^0
             u_2^{\rho_2}(\rho_1{\bf p},\lambda_2)         \,,                                                                                                                               
                               \end{eqnarray} 
                which follow directly from the Dirac equations for $\bar u_1^{\rho_1}(\rho_2{\bf p},\lambda_1)$ and $u_2^{\rho_2}(\rho_1{\bf p},\lambda_2)$. \\              
In the following subsections we present, for each meson, the expressions for the 4CSE wave functions in the partial-wave tensor basis. We always work in the meson rest frame where $P=(\mu,{\bf 0})$.

 %%%%%%%%%%%%%%%%%%%%%%%%%%%%%%%%%%%%%%%%%%%%%%%%%%%%%%%%%
\begin{widetext}
\subsubsection{Pseudoscalar mesons}
 For the extraction of the $P$ and $S$ wave components from the CST wave function for a pseudoscalar meson we have to distinguish between the two cases where the $\rho$-spins of the incoming and outgoing quarks are the same ($\rho_1=\rho_2$) or the opposite ($\rho_1=-\rho_2$). Furthermore, we have to distinguish whether quark 1 or quark 2 is on mass shell.
For $\rho_1=\rho_2$, and quark 1 on mass shell, we obtain for the spinor matrix elements of the vertex function, after a short calculation, the expression
\begin{eqnarray}
 \bar u_1^{\rho_1}(\rho_1{\bf p},\lambda_1)\Gamma^{p}_{1\rho_1}(p)u_2^{\rho_1}(\rho_1{\bf p},\lambda_2)&=&N_{1p}N_{2p} \left\lbrace 
 \rho_1G^{p\rho_1}_{1,1}  \left[\tilde p_2-\tilde p_1\right] + G^{p\rho_1}_{2,1} \frac{\rho_1(E_{1 p}-E_{2 p})-\mu}{2m_{2}}  \left[\tilde p_2+\tilde p_1\right] \right\rbrace 
 \chi^\dag_{1}\boldsymbol\sigma\cdot \hat {\bf p} \chi_{2}\,,\nonumber\\
\end{eqnarray}
where we have introduced $\tilde p_i=p/(E_{ip}+m_i)$ and the shorthand $\chi_{i}\equiv \chi_{\lambda_i}$. 
The 
analogous expression when quark 2 is on mass shell reads
\begin{eqnarray}
 \bar u_1^{\rho_2}(\rho_2{\bf p},\lambda_1)\Gamma^{p}_{2\rho_2}(p)u_2^{\rho_2}(\rho_2{\bf p},\lambda_2)&=&N_{1p}N_{2p} \left\lbrace 
 \rho_2G^{p\rho_2}_{1,2}  \left[\tilde p_2-\tilde p_1\right] + G^{p\rho_2}_{3,2} \frac{\rho_2(E_{1 p}-E_{2 p})-\mu}{2m_{1}}  \left[\tilde p_2+\tilde p_1\right] \right\rbrace 
 \chi^\dag_{1}\boldsymbol\sigma\cdot \hat {\bf p}\chi_{2}\,.\nonumber\\
\end{eqnarray}
The CST wave functions, as defined in Eq.~(\ref{eq:CSTwfs}) and for quark 1 on shell then become 
\begin{eqnarray}
 \Psi ^{p\rho_1\rho_1}_{1,\lambda_1\lambda_2}({\bf p})
 &=& 
 -\frac{1}{2}\mathcal N_{12p}  \left\lbrace 
 \frac{G^{p\rho_1}_{1,1} }{E_{1 p}-E_{2 p}-\rho_{1}\mu}  \left[\tilde p_2-\tilde p_1\right] + \frac{G^{p\rho_1}_{2,1}}{2m_{2}}  \left[\tilde p_2+\tilde p_1\right] \right\rbrace 
 \chi^\dag_{1}\boldsymbol\sigma\cdot \hat {\bf p} \chi_{2}= \psi^{p\rho_1}_{P, 1}( p) \chi^\dag_{1}\boldsymbol\sigma \cdot \hat {\bf p} \chi_{2}
 \,
\end{eqnarray}
where $\mathcal N_{12p}=\frac{\sqrt{E_{1 p}+m_1} \sqrt{E_{2 p}+m_2}}{\sqrt{E_{1 p}E_{2 p}}}$. For quark 2 on shell the 
analogous expression reads
\begin{eqnarray}
 \Psi ^{p\rho_2\rho_2}_{2,\lambda_1\lambda_2}({\bf p})
 &=& 
 \frac{1}{2}\mathcal N_{12p}  \left\lbrace 
 \frac{G^{p\rho_2}_{1,2} }{E_{1 p}-E_{2 p}-\rho_{2}\mu}  \left[\tilde p_2-\tilde p_1\right] + \frac{G^{p\rho_2}_{3,2}}{2m_{1}}  \left[\tilde p_2+\tilde p_1\right] \right\rbrace 
 \chi^\dag_{1}\boldsymbol\sigma\cdot \hat {\bf p} \chi_{2}= \psi^{p\rho_2}_{P, 2}( p) \chi^\dag_{1}\boldsymbol\sigma \cdot \hat {\bf p} \chi_{2}
 \,.
\end{eqnarray}
From these expressions one can read off the $P$-waves $\psi^{p\rho_i}_{P, i}( p) $ when quark $i$ is on mass shell with positive ($\rho_i=+$) or negative ($\rho_i=-$) energy. For the case of the 1CSE we identify $\psi^{+}_{1}(p)\equiv\psi^{p+}_{P, 1}(p) $ and $ K_1^+ (\hat {\bf p})\equiv \boldsymbol\sigma \cdot \hat {\bf p}$.
\\
For $\rho_1=-\rho_2$, and quark 1 or quark 2 on-shell we obtain the expressions
\begin{eqnarray}
\bar u_1^{\rho_1}(\rho_1{\bf p},\lambda_1)\Gamma^{p}_{1\rho_1}(p)u_2^{-\rho_1}(\rho_1{\bf p},\lambda_2)&=&N_{1p}N_{2p}\left\lbrace 
 \rho_1G^{p\rho_1}_{1,1}  \left[1+\tilde p_1\tilde p_2\right] -G^{p\rho_1}_{2,1} \frac{-\rho_1(E_{1 p}+E_{2 p})+\mu}{2m_{2}}  \left[1-\tilde p_1\tilde p_2\right] \right\rbrace 
 \chi^\dag_{1} \mathbf{1} \chi_{2}\, ,\nonumber\\
\\
 \bar u_1^{-\rho_2}(\rho_2{\bf p},\lambda_1)\Gamma^{p}_{2\rho_2}(p)u_2^{\rho_2}(\rho_2{\bf p},\lambda_2)&=&N_{1p}N_{2p}\left\lbrace 
 -\rho_2G^{p\rho_2}_{1,2}  \left[1+\tilde p_1\tilde p_2\right] + G^{p\rho_2}_{3,2} \frac{-\rho_2(E_{1 p}+E_{2 p})-\mu}{2m_{1}}  \left[1-\tilde p_1\tilde p_2\right] \right\rbrace 
 \chi^\dag_{1} \mathbf 1\chi_{2}\, .\nonumber\\
\end{eqnarray}
The corresponding wave functions then read
\begin{eqnarray}
 \Psi ^{p\rho_1-\rho_1}_{1,\lambda_1\lambda_2}({\bf p})
  &=& -\frac{1}{2}\mathcal N_{12p} \left\lbrace \frac{G^{p\rho_1}_{1,1}}{E_{1 p}+ E_{2 p}-\rho_1\mu}  \left[1+\tilde p_1\tilde p_2\right]  + \frac{G^{p\rho_1}_{2,1}}{2m_{2}}  \left[1-\tilde p_1\tilde p_2\right]
  \right\rbrace\chi^\dag_{1}\mathbf 1\chi_{2} = \psi_{S,1}^{p\rho_1}( p) \chi^\dag_{1}\mathbf 1\chi_{2} \, , \\
 \Psi ^{p-\rho_2\rho_2}_{2,\lambda_1\lambda_2}({\bf p})
  &=& \frac{1}{2}\mathcal N_{12p} \left\lbrace \frac{G^{p\rho_2}_{1,2}}{E_{1 p}+ E_{2 p}+\rho_2\mu}  \left[1+\tilde p_1\tilde p_2\right]  + \frac{G^{p\rho_2}_{3,2}}{2m_{1}}  \left[1-\tilde p_1\tilde p_2\right]
  \right\rbrace\chi^\dag_{1}\mathbf 1\chi_{2} = \psi_{S,2}^{p\rho_2}( p) \chi^\dag_{1}\mathbf 1\chi_{2}\,,
\end{eqnarray} 
from which one can read off the $S$-waves $\psi^{p\rho_i}_{S, i}( p) $.
For the case of the 1CSE we identify $\psi^{-}_{1}(p)\equiv \psi^{p-}_{S, 1}(p)$ and $ K_1^- (\hat {\bf p})\equiv{\bf 1} $. 
\\
The 1CSE wave function components are normalized as
\begin{eqnarray}
 \int_0^\infty \mathrm dp \,p^2 \left[\left(\psi_{S,1}^{p-}(p)\right)^2+ \left(\psi_{P,1}^{p+}(p)\right)^2\right]=1\,.
 \label{eq:pnorm}
\end{eqnarray}

\subsubsection{Scalar mesons}
 The treatment of the scalar mesons is very similar to the previous one of pseudoscalar mesons. 
For $\rho_2=\rho_1$, and quark 1 or quark 2 on mass shell the CST wave functions read 
\begin{eqnarray}
 \Psi ^{s\rho_1\rho_1}_{1,\lambda_1\lambda_2}({\bf p})
 &=& -\frac{1}{2}\mathcal N_{12p}   \left\lbrace \frac{G^{s\rho_1}_{1,1}}{E_{1 p}-E_{2 p}-\rho_1\mu}
   \left[1-\tilde p_1\tilde p_2\right] - \frac{G^{s\rho_1}_{2,1}}{2m_{2}}  \left[1+\tilde p_1\tilde p_2\right] \right\rbrace 
 \chi^\dag_{1}\mathbf 1\chi_{2} 
 = \psi^{s\rho_1}_{S,1}(p)\chi^\dag_{1}\mathbf 1\chi_{2}\, , \\
 \Psi ^{s\rho_2\rho_2}_{2,\lambda_1\lambda_2}({\bf p})
 &=& \frac{1}{2}\mathcal N_{12p}   \left\lbrace \frac{G^{s\rho_2}_{1,2}}{E_{1 p}-E_{2 p}-\rho_2\mu}
   \left[1-\tilde p_1\tilde p_2\right] + \frac{G^{s\rho_2}_{3,2}}{2m_{1}}  \left[1+\tilde p_1\tilde p_2\right] \right\rbrace 
 \chi^\dag_{1}\mathbf 1\chi_{2}
 =\psi^{s\rho_2}_{S,2}(p)\chi^\dag_{1}\mathbf 1\chi_{2}\, .
 \,
\end{eqnarray}
For the 1CSE we identify $\psi^{+}_{1}(p)\equiv \psi^{s+}_{S, 1}(p)$ and $ K_1^+ (\hat {\bf p})\equiv{\bf 1} $.\\
For $\rho_1=-\rho_2$, and quark 1 or quark 2 on mass shell we have 
\begin{eqnarray}
 \Psi ^{s\rho_1-\rho_1}_{1,\lambda_1\lambda_2}({\bf p})
  &=& \frac{1}{2}\mathcal N_{12p}  \left\lbrace \frac{G^{s\rho_1}_{1,1}}{E_{1 p}+ E_{2 p}-\rho_1\mu}  \left[\tilde p_2+\tilde p_1\right] - \frac{G^{s\rho_1}_{2,1}}{2m_{2}}  \left[\tilde p_2-\tilde p_1\right]
  \right\rbrace\chi^\dag_{1}\boldsymbol\sigma\cdot \hat{\bf p} \chi_{2}  = \psi_{P,1}^{s\rho_1}(p)\chi^\dag_{1}\boldsymbol\sigma\cdot \hat{\bf p} \chi_{2} \, , \\
 \Psi ^{s-\rho_2\rho_2}_{2,\lambda_1\lambda_2}({\bf p})
  &=& \frac{1}{2}\mathcal N_{12p}  \left\lbrace \frac{G^{s\rho_2}_{1,2}}{E_{1 p}+ E_{2 p}+\rho_2\mu}  \left[\tilde p_2+\tilde p_1\right] + \frac{G^{s\rho_2}_{3,2}}{2m_{1}}  \left[\tilde p_2-\tilde p_1\right]
  \right\rbrace\chi^\dag_{1}\boldsymbol\sigma\cdot \hat{\bf p} \chi_{2}  =\psi_{P,2}^{s\rho_2}(p)\chi^\dag_{1}\boldsymbol\sigma\cdot \hat{\bf p} \chi_{2} \, .
\end{eqnarray}
For the 1CSE we identify $\psi^{-}_{1}(p)\equiv\psi^{s-}_{P, 1}(p) $ and $ K_1^- (\hat {\bf p})\equiv \boldsymbol\sigma \cdot \hat {\bf p}$.
\\
The 1CSE wave function components are normalized as
\begin{eqnarray}
 \int_0^\infty \mathrm dp \,p^2 \left[\left(\psi_{S,1}^{s+}(p)\right)^2+ \left(\psi_{P,1}^{s-}(p)\right)^2\right]=1\,.
 \label{eq:snorm}
\end{eqnarray}

%%%%%%%%%%%%%%%%%%%%%%%%%%%%%%%%%%%%%%%%%%%%%%%%%%%%%
\subsubsection{Vector mesons}
For vector mesons, the $S$, $P_s$, $P_t$, and $D$ wave components are extracted from the CST wave function in a similar way as in the previous spin-0 meson cases. For $\rho_1=\rho_2$, and quark 1 or quark 2 on mass shell 
the CST wave functions read
\begin{eqnarray}
 \Psi^{v\rho_1\rho_1}_{1,\lambda_1\lambda_2}({\bf p})&=&
 -\frac{1}{2}\mathcal N_{12p}
 \left\lbrace -
 \left[ p\left(-\frac{(1+\tilde p_1 \tilde p_2)G_{6,1}^{v\rho_1}}{2m_{2}} +\frac{\rho_1(1-\tilde p_1 \tilde p_2)G_{5,1}^{v\rho_1}}{\rho_1( E_{1p}- E_{2p})-\mu} \right)\right.\right.\nonumber\\&&+\left.\left.\left(-\frac{\rho_1(\tilde p_1-\tilde p_2)G_{2,1}^{v\rho_1}}{2m_{2}} +\frac{(\tilde p_1+\tilde p_2)G_{1,1}^{v\rho_1}}{\rho_1( E_{1p}- E_{2p})-\mu} \right)  \right]\boldsymbol\xi\cdot \hat{{\bf p}}\chi_{1}^{\dag}\mathbf 1\chi_{2}\right.\nonumber\\&&+\left.\left[-\frac{\rho_1(\tilde p_1+\tilde p_2)G_{2,1}^{v\rho_1}}{2m_{2}} + \frac{(\tilde p_1-\tilde p_2)G_{1,1}^{v\rho_1}}{\rho_1( E_{1p}- E_{2p})-\mu}\right] \chi_{1}^{\dag}\left(\boldsymbol\sigma \cdot \boldsymbol\xi \boldsymbol\sigma\cdot\hat{{\bf p}}-\boldsymbol\xi\cdot \hat{{\bf p}}\right)\chi_{2}
 \right\rbrace  \nonumber\\
 &=&\sqrt 3\psi_{P_s,1}^{v\rho_1}(p)  \boldsymbol\xi \cdot \hat{{\bf p}}\chi_{1}^{\dag}\mathbf 1  \chi_{2}+\sqrt{\frac{3}{2}}\psi_{P_t,1}^{v\rho_1}(p) \chi_{1}^{\dag}\left(\boldsymbol\sigma\cdot \boldsymbol\xi\boldsymbol\sigma\cdot \hat{{\bf p}}- \boldsymbol\xi \cdot \hat{{\bf p}}  \right)\chi_{2} \, ,
 \label{eq:PsiVrhopar1} \\
 \Psi^{v\rho_2\rho_2}_{2,\lambda_1\lambda_2}({\bf p})&=&
 \frac{1}{2}\mathcal N_{12p}
 \left\lbrace -
 \left[ p\left(\frac{(1+\tilde p_1 \tilde p_2)G_{7,2}^{v\rho_2}}{2m_1} +\frac{\rho_2(1-\tilde p_1 \tilde p_2)G_{5,2}^{v\rho_2}}{\rho_2( E_{1p}- E_{2p})-\mu} \right)\right.\right.\nonumber\\&&+\left.\left.\left(-\frac{\rho_2(\tilde p_1-\tilde p_2)G_{3,2}^{v\rho_2}}{2m_1} +\frac{(\tilde p_1+\tilde p_2)G_{1,2}^{v\rho_2}}{\rho_2( E_{1p}- E_{2p})-\mu} \right)  \right]\boldsymbol\xi\cdot \hat{{\bf p}}\chi_{1}^{\dag}\mathbf 1\chi_{2}\right.\nonumber\\&&+\left.\left[-\frac{\rho_2(\tilde p_1+\tilde p_2)G_{3,2}^{v\rho_2}}{2m_1} + \frac{(\tilde p_1-\tilde p_2)G_{1,2}^{v\rho_2}}{\rho_2( E_{1p}- E_{2p})-\mu}\right] \chi_{1}^{\dag}\left(\boldsymbol\sigma \cdot \boldsymbol\xi \boldsymbol\sigma\cdot\hat{{\bf p}}-\boldsymbol\xi\cdot \hat{{\bf p}}\right) \chi_{2}
 \right\rbrace  \nonumber\\
 &=&\sqrt 3\psi_{P_s,2}^{v\rho_2}(p)  \boldsymbol\xi \cdot \hat{{\bf p}}\chi_{1}^{\dag} \mathbf 1 \chi_{2}+\sqrt{\frac{3}{2}}\psi_{P_t,2}^{v\rho_2} (p)\chi_{1}^{\dag}\left(\boldsymbol\sigma\cdot \boldsymbol\xi\boldsymbol\sigma\cdot \hat{{\bf p}}- \boldsymbol\xi \cdot \hat{{\bf p}}  \right)\chi_{2} \, .
 \label{eq:PsiVrhopar2}
\end{eqnarray}
From these expressions we can read off the spin-singlet and spin-triplet $P$ waves $\psi_{P_s,i}^{v\rho_i}(p)$ and $\psi_{P_t,i}^{v\rho_i}(p)$, respectively. For the 1CSE case we identify $\psi^+_1(p)\equiv \psi_{P_s,1}^{v+}(p)$, $\psi^+_2(p)\equiv \psi_{P_t,1}^{v+}(p)$, $K_1^+ (\hat {\bf p}) \equiv \sqrt 3 \boldsymbol\xi \cdot \hat{{\bf p}}$, and $K_2^+ (\hat {\bf p}) \equiv \sqrt{\frac{3}{2}}\left(\boldsymbol\sigma\cdot \boldsymbol\xi\boldsymbol\sigma\cdot \hat{{\bf p}}- \boldsymbol\xi \cdot \hat{{\bf p}}  \right)$. 
\\
For $\rho_1=-\rho_2$ and quark 1 or quark 2 on mass shell 
the corresponding wave functions are given by 
\begin{eqnarray}\Psi^{v\rho_1-\rho_1}_{1,\lambda_1\lambda_2}({\bf p})
 &=&
\frac{1}{2}\mathcal N_{12p}
 \left\lbrace \left[\frac{\rho_1(3-\tilde  p_1\tilde  p_2)G_{2,1}^{v\rho_1}}{6m_2} -\frac{(3+\tilde  p_1\tilde  p_2)G_{1,1}^{v\rho_1}}{ -3\rho_1(E_{1 p}+E_{2 p})+3\mu}\right.\right.\nonumber\\&&
 +\left.\left.\frac{(\tilde  p_1 - \tilde p_2)G_{6,1}^{v\rho_1}}{2m_2} +\frac{\rho_1(\tilde  p_1 +\tilde  p_2)G_{5,1}^{v\rho_1}}{ -\rho_1(E_{1 p}+E_{2 p})+\mu}
 \right] \chi_{1}^{\dag}\boldsymbol\sigma \cdot \boldsymbol\xi \chi_{2}\right.\nonumber\\&&
 +\left.
 \frac{ p}{3}\left[\frac{2\tilde  p_1\tilde  p_2}{ p} \left(\frac{\rho_1G_{2,1}^{v\rho_1}}{2m_2}+ \frac{G_{1,1}^{v\rho_1}}{ -\rho_1(E_{1 p}+E_{2 p})+\mu}\right) \right.\right.\nonumber\\&&
 +\left.\left.\frac{(\tilde  p_1 - \tilde p_2)G_{6,1}^{v\rho_1}}{2m_2} +\frac{\rho_1(\tilde  p_1 + \tilde p_2)G_{5,1}^{v\rho_1}}{ -\rho_1(E_{1 p}+E_{2 p})+\mu} \right] \chi_{1}^{\dag} (3\boldsymbol\xi\cdot \hat{{\bf p}}\boldsymbol\sigma\cdot \hat{{\bf p}} -\boldsymbol\sigma \cdot \boldsymbol\xi )\chi_{2}
 \right\rbrace \, \nonumber\\
 &=&\psi_{S,1}^{v\rho_1} (p)\chi_{1}^{\dag} \boldsymbol\sigma \cdot \boldsymbol\xi \chi_{2}+\frac{1}{\sqrt 2}\psi_{D,1}^{v\rho_1}(p) \chi_{1}^{\dag}\left(3 \boldsymbol\xi\cdot \hat{\bf p} \boldsymbol\sigma\cdot \hat{\bf  p}- \boldsymbol\sigma\cdot\boldsymbol\xi \right)\chi_{2}\,, 
 \label{eq:PsiVrhoantip1} \\
\Psi^{v-\rho_2\rho_2}_{2,\lambda_1\lambda_2}({\bf p})
 &=&
\frac{1}{2}\mathcal N_{12p}
 \left\lbrace \left[\frac{\rho_2(3-\tilde  p_1\tilde  p_2)G_{3,2}^{v\rho_2}}{6m_1} -\frac{(3+\tilde  p_1\tilde  p_2)G_{1,2}^{v\rho_2}}{ -3\rho_2(E_{1 p}+E_{2 p})-3\mu}\right.\right.\nonumber\\&&
 +\left.\left.\frac{-(\tilde  p_1 - \tilde p_2)G_{7,2}^{v\rho_2}}{2m_1} +\frac{\rho_2(\tilde  p_1 +\tilde  p_2)G_{5,2}^{v\rho_2}}{ -\rho_2(E_{1 p}+E_{2 p})-\mu}
 \right] \chi_{1}^{\dag}\boldsymbol\sigma \cdot \boldsymbol\xi \chi_{2}\right.\nonumber\\&&
 +\left.
 \frac{ p}{3}\left[\frac{2\tilde  p_1\tilde  p_2}{ p} \left(\frac{\rho_2G_{3,2}^{v\rho_2}}{2m_1}+ \frac{G_{1,2}^{v\rho_2}}{ -\rho_2(E_{1 p}+E_{2 p})-\mu}\right) \right.\right.\nonumber\\&&
 +\left.\left.\frac{-(\tilde  p_1 - \tilde p_2)G_{7,2}^{v\rho_2}}{2m_1} +\frac{\rho_2(\tilde  p_1 + \tilde p_2)G_{5,2}^{v\rho_2}}{ -\rho_2(E_{1 p}+E_{2 p})-\mu} \right] \chi_{1}^{\dag} (3\boldsymbol\xi\cdot \hat{{\bf p}}\boldsymbol\sigma\cdot \hat{{\bf p}} -\boldsymbol\sigma \cdot \boldsymbol\xi )\chi_{2}
 \right\rbrace \, \nonumber\\
 &=&\psi_{S,2}^{v\rho_2} (p)\chi_{1}^{\dag} \boldsymbol\sigma \cdot \boldsymbol\xi \chi_{2}+\frac{1}{\sqrt 2}\psi_{D,2}^{v\rho_2}(p) \chi_{1}^{\dag}\left(3 \boldsymbol\xi\cdot \hat{\bf p} \boldsymbol\sigma\cdot \hat{\bf  p}- \boldsymbol\sigma\cdot\boldsymbol\xi \right)\chi_{2}\, .
 \label{eq:PsiVrhoantip2}
\end{eqnarray}
From these expressions we can read off the $S$ and $D$ waves $\psi_{S,i}^{v\rho_i}(p)$ and $\psi_{D,i}^{v\rho_i}(p)$, respectively. For the 1CSE we identify $\psi^-_1(p)\equiv \psi_{S,1}^{v-}(p)$, $\psi^-_2(p)\equiv \psi_{D,1}^{v-}(p)$, $K_1^- (\hat {\bf p}) \equiv \boldsymbol\sigma \cdot \boldsymbol\xi $, and $K_2^- (\hat {\bf p}) \equiv \frac{1}{\sqrt 2}\left(3 \boldsymbol\xi\cdot \hat{\bf p} \boldsymbol\sigma\cdot \hat{\bf p}- \boldsymbol\sigma\cdot\boldsymbol\xi \right)$. 
\\
The 1CSE wave function components are normalized as
\begin{eqnarray}
 \int_0^\infty \mathrm dp\, p^2 \left[\left(\psi_{S,1}^{v-}(p)\right)^2+ \left(\psi_{P_s,1}^{v+}(p)\right)^2+ \left(\psi_{P_t,1}^{v+}(p)\right)^2+ \left(\psi_{D,1}^{v-}(p)\right)^2\right]=1\,.
 \label{eq:vnorm}
\end{eqnarray}

%%%%%%%%%%%%%%%%%%%%%%%%%%%%%%%%%%%%%%%%%%%%%%%%%?
\subsubsection{Axial-vector mesons}
The treatment of the axial-vector mesons is very similar to the previous one of vector mesons. 
For $\rho_1=\rho_2$,  and quark 1 or quark 2 on mass shell the CST wave functions read
\begin{eqnarray}\Psi^{a\rho_1\rho_1}_{1,\lambda_1\lambda_2}({\bf p})
 &=&
\frac{1}{2}\mathcal N_{12p}
 \left\lbrace \left[-\frac{\rho_1(3+\tilde  p_1\tilde  p_2)G_{2,1}^{a\rho_1}}{6m_2} -\frac{-(3-\tilde  p_1\tilde  p_2)G_{1,1}^{a\rho_1}}{ 3\rho_1(E_{1 p}-E_{2 p})-3\mu}\right.\right.\nonumber\\&&
 -\left.\left.\frac{-(\tilde  p_1 + \tilde p_2)
G_{6,1}^{a\rho_1}}{2m_2} +\frac{-\rho_1(\tilde  p_1 -\tilde  p_2)G_{5,1}^{a\rho_1}}{\rho_1(E_{1 p}-E_{2 p})-\mu}
 \right] \chi_{1}^{\dag}\boldsymbol\sigma \cdot \boldsymbol\xi \chi_{2}\right.\nonumber\\&&
 +\left.
 \frac{ p}{3}\left[-\frac{2\tilde  p_1\tilde  p_2}{ p} \left(-\frac{\rho_1G_{2,1}^{a\rho_1}}{2m_2}+ \frac{-G_{1,1}^{a\rho_1}}{ \rho_1(E_{1 p}-E_{2 p})-\mu}\right) \right.\right.\nonumber\\&&
 -\left.\left.\frac{-(\tilde  p_1 + \tilde p_2)G_{6,1}^{a\rho_1}}{2m_2} +\frac{-\rho_1(\tilde  p_1 - \tilde p_2)G_{5,1}^{a\rho_1}}{ \rho_1(E_{1 p}-E_{2 p})-\mu} \right] \chi_{1}^{\dag} (3\boldsymbol\xi\cdot \hat{{\bf p}}\boldsymbol\sigma\cdot \hat{{\bf p}} -\boldsymbol\sigma \cdot \boldsymbol\xi )\chi_{2}
 \right\rbrace \, \nonumber\\
 &=&\psi_{S,1}^{a\rho_1}(p) \chi_{1}^{\dag} \boldsymbol\sigma \cdot \boldsymbol\xi \chi_{2}+\frac{1}{\sqrt 2}\psi_{D,1}^{a\rho_1}(p) \chi_{1}^{\dag}\left(3 \boldsymbol\xi\cdot \hat{\bf  p} \boldsymbol\sigma\cdot \hat{\bf p}- \boldsymbol\sigma\cdot\boldsymbol\xi \right)\chi_{2}\, ,
 \label{eq:PsiVrhoantip1} \\
\Psi^{a\rho_2\rho_2}_{2,\lambda_1\lambda_2}({\bf p})
 &=&
\frac{1}{2}\mathcal N_{12p}
 \left\lbrace \left[-\frac{\rho_2(3+\tilde  p_1\tilde  p_2)G_{3,2}^{a\rho_2}}{6m_1} -\frac{(3-\tilde  p_1\tilde  p_2)G_{1,2}^{a\rho_2}}{ 3\rho_2(E_{1 p}-E_{2 p})-3\mu}\right.\right.\nonumber\\&&
 -\left.\left.\frac{(\tilde  p_1 + \tilde p_2)
G_{7,2}^{a\rho_2}}{2m_1} +\frac{\rho_2(\tilde  p_1 -\tilde  p_2)G_{5,2}^{a\rho_2}}{\rho_2(E_{1 p}-E_{2 p})-\mu}
 \right] \chi_{1}^{\dag}\boldsymbol\sigma \cdot \boldsymbol\xi \chi_{2}\right.\nonumber\\&&
 +\left.
 \frac{ p}{3}\left[-\frac{2\tilde  p_1\tilde  p_2}{ p} \left(-\frac{\rho_2G_{3,2}^{a\rho_2}}{2m_1}+ \frac{G_{1,2}^{a\rho_2}}{ \rho_2(E_{1 p}-E_{2 p})-\mu}\right) \right.\right.\nonumber\\&&
 -\left.\left.\frac{(\tilde  p_1 + \tilde p_2)G_{7,2}^{a\rho_2}}{2m_1} +\frac{\rho_2(\tilde  p_1 - \tilde p_2)G_{5,2}^{a\rho_2}}{ \rho_2(E_{1 p}-E_{2 p})-\mu} \right] \chi_{1}^{\dag} (3\boldsymbol\xi\cdot \hat{{\bf p}}\boldsymbol\sigma\cdot \hat{{\bf p}} -\boldsymbol\sigma \cdot \boldsymbol\xi )\chi_{2}
 \right\rbrace \, \nonumber\\
 &=&\psi_{S,2}^{a\rho_2}(p) \chi_{1}^{\dag} \boldsymbol\sigma \cdot \boldsymbol\xi \chi_{2}+\frac{1}{\sqrt 2}\psi_{D,2}^{a\rho_2}(p) \chi_{1}^{\dag}\left(3 \boldsymbol\xi\cdot \hat{\bf  p} \boldsymbol\sigma\cdot \hat{\bf p}- \boldsymbol\sigma\cdot\boldsymbol\xi \right)\chi_{2}\, .
 \label{eq:PsiVrhoantip}
\end{eqnarray}
For the 1CSE we identify $\psi^+_1(p)\equiv \psi_{S,1}^{a+}(p)$ and $\psi^+_2(p)\equiv \psi_{D,1}^{a+}(p)$, and $K_1^+ (\hat {\bf p}) \equiv \boldsymbol\sigma \cdot \boldsymbol\xi $ and $K_2^+ (\hat {\bf p}) \equiv \frac{1}{\sqrt 2}\left(3 \boldsymbol\xi\cdot \hat{\bf p} \boldsymbol\sigma\cdot \hat{\bf p}- \boldsymbol\sigma\cdot\boldsymbol\xi \right)$. 
\\
For $\rho_1=-\rho_2$, and quark 1 or quark 2 on mass shell they read
\begin{eqnarray}
 \Psi^{a\rho_1-\rho_1}_{1,\lambda_1\lambda_2}({\bf p})&=& \frac{1}{2}
 \mathcal N_{1 2p}
 \left\lbrace -
 \left[  p\left(-\frac{(1-\tilde  p_1 \tilde  p_2)G_{6,1}^{a\rho_1}}{2m_2}  +\frac{\rho_1(1+\tilde  p_1 \tilde  p_2)  G_{5,1}^{a\rho_1}}{-\rho_1( E_{1 p}+ E_{2 p})+\mu}\right) \right.\right.\nonumber\\&&-\left.\left.\rho_1(\tilde  p_1+\tilde  p_2)\frac{G_{2,1}^{a\rho_1}}{2m_2}  + \frac{(\tilde  p_1-\tilde  p_2)G_{1,1}^{a\rho_1}}{-\rho_1( E_{1 p}+ E_{2 p})+\mu} \right]\boldsymbol\xi\cdot \hat{\bf p}\chi_{1}^{\dag}\mathbf 1\chi_{2}\right.\nonumber\\&&+\left.\left[\rho_1(\tilde  p_2-\tilde  p_1)\frac{G_{2,1}^{a\rho_1}}{2m_2}  + \frac{(\tilde  p_1+\tilde  p_2) G_{1,1}^{a\rho_1}}{-\rho_1( E_{1 p}+ E_{2 p})+\mu}\right] \chi_{1}^{\dag}\left(\boldsymbol\sigma \cdot \boldsymbol\xi \boldsymbol\sigma\cdot\hat{{\bf p}} -\boldsymbol\xi\cdot \hat{\bf p}\right) \chi_{2}
 \right\rbrace 
 \nonumber\\
 &=&\sqrt 3\psi_{P_s,1}^{a\rho_1}(p)  \boldsymbol\xi \cdot \hat{{\bf p}}\chi_{1}^{\dag} \mathbf 1 \chi_{2}+\sqrt{\frac{3}{2}}\psi_{P_t,1}^{a\rho_1} (p)\chi_{1}^{\dag}\left(\boldsymbol\sigma\cdot \boldsymbol\xi\boldsymbol\sigma\cdot \hat{{\bf p}}- \boldsymbol\xi \cdot \hat{{\bf p}}  \right)\chi_{2} \,,
 \label{eq:GammaArhoantip1} \\
 \Psi^{a-\rho_2\rho_2}_{2,\lambda_1\lambda_2}({\bf p})&=& -\frac{1}{2}
 \mathcal N_{1 2p}
 \left\lbrace -
 \left[  p\left(-\frac{(1-\tilde  p_1 \tilde  p_2)G_{7,2}^{a\rho_2}}{2m_1}  +\frac{\rho_2(1+\tilde  p_1 \tilde  p_2)  G_{5,2}^{a\rho_2}}{-\rho_2( E_{1 p}+ E_{2 p})-\mu}\right) \right.\right.\nonumber\\&&\left.\left.+\rho_2(\tilde  p_1+\tilde  p_2)\frac{G_{3,2}^{a\rho_2}}{2m_1}  + \frac{(\tilde  p_1-\tilde  p_2)G_{1,2}^{a\rho_2}}{-\rho_2( E_{1 p}+ E_{2 p})-\mu} \right]\boldsymbol\xi\cdot \hat{\bf p}\chi_{1}^{\dag}\mathbf 1\chi_{2}\right.\nonumber\\&&+\left.\left[-\rho_2(\tilde  p_2-\tilde  p_1)\frac{G_{3,2}^{a\rho_2}}{2m_1}  + \frac{(\tilde  p_1+\tilde  p_2) G_{1,2}^{a\rho_2}}{-\rho_2( E_{1 p}+ E_{2 p})-\mu}\right] \chi_{1}^{\dag}\left(\boldsymbol\sigma \cdot \boldsymbol\xi \boldsymbol\sigma\cdot\hat{{\bf p}} -\boldsymbol\xi\cdot \hat{\bf p}\right) \chi_{2}
 \right\rbrace 
 \nonumber\\
 &=&\sqrt 3\psi_{P_s,2}^{a\rho_2} (p) \boldsymbol\xi \cdot \hat{{\bf p}}\chi_{1}^{\dag} \mathbf 1\chi_{2}+\sqrt{\frac{3}{2}}\psi_{P_t,2}^{a\rho_2} (p)\chi_{1}^{\dag}\left(\boldsymbol\sigma\cdot \boldsymbol\xi\boldsymbol\sigma\cdot \hat{{\bf p}}- \boldsymbol\xi \cdot \hat{{\bf p}}  \right)\chi_{2} \, .
\label{eq:GammaArhoantip2}
\end{eqnarray} 
For the 1CSE we identify $\psi^-_1(p)\equiv \psi_{P_s,1}^{a-}(p)$ and $\psi^-_2(p)\equiv \psi_{P_t,1}^{a-}(p)$, and $K_1^- (\hat {\bf p}) \equiv \sqrt 3 \boldsymbol\xi \cdot \hat{{\bf p}}$ and $K_2^- (\hat {\bf p}) \equiv \sqrt{\frac{3}{2}}\left(\boldsymbol\sigma\cdot \boldsymbol\xi\boldsymbol\sigma\cdot \hat{{\bf p}}- \boldsymbol\xi \cdot \hat{{\bf p}}  \right)$. \\
The 1CSE wave function components are normalized as
\begin{eqnarray}
 \int_0^\infty \mathrm dp\, p^2 \left[\left(\psi_{S,1}^{a+}(p)\right)^2+ \left(\psi_{P_s,1}^{a-}(p)\right)^2+ \left(\psi_{P_t,1}^{a-}(p)\right)^2+ \left(\psi_{D,1}^{a+}(p)\right)^2\right]=1\,.
\label{eq:wfnorm}
\end{eqnarray}

\end{widetext}
%%%%%%%%%%%%%%%%%%%%%%%%%%%%%%%%%%%%%%%%%%%%%%%%%%%%%%%%%%

\section{Vertex spinor matrix elements $M_i^{K,\rho\rho'}$} 
\label{app:B} \vspace{0.1in}
Here we give the explicit expressions of the $M_i^{K,\rho\rho'}$ functions as defined in Eqs.~(\ref{eq:thetaO}) and (\ref{eq:thetaO2}), for each Lorentz structure $K$ of the interaction kernel: $K=s$ (scalar), $p$ (pseudoscalar), and $v$ (vector).

 \begin{align}
\Theta^s_i={\bf 1}: \hspace{1cm} &\nonumber\\
M_i^{S,++}({\bf p},{\bf k}) & =1 - \tilp_i \tilk_i \cp\, \ck \\
M_i^{S,+-}({\bf p},{\bf k}) & =  -\tilk_i\ck -\tilp_i \cp \\
 M_i^{S,-+}({\bf p},{\bf k}) & = -\tilp_i  \cp -\tilk_i \ck \\
 M_i^{S,--}({\bf p},{\bf k}) & = \tilp_i\tilk_i \cp\, \ck -1 \, ,\\
 \vspace{2mm}\nonumber\\
\Theta^p_i=\gamma^5:\hspace{1cm} &\nonumber\\
M_i^{P,++}({\bf p},{\bf k}) & =-1 - \tilp_i \, \tilk_i \,\cp\, \ck \\
M_i^{P,+-}({\bf p},{\bf k}) & = \tilk_i\, \ck -\tilp_i \, \cp \\
 M_i^{P,-+}({\bf p},{\bf k}) & = -\tilk_i  \,\ck +\tilp_i \, \cp \\
 M_i^{P,--}({\bf p},{\bf k}) & = -1 - \tilp_i \, \tilk_i \, \cp\, \, \ck \, ,
\end{align}
\\

\begin{align}
\Theta^{v0}_i =\gamma^0:\hspace{1cm}& \nonumber\\
M_i^{V0,++}({\bf p},{\bf k}) & = 1+ \tilp_i \, \tilk_i \,\cp\, \ck \\
M_i^{V0,+-}({\bf p},{\bf k}) & =-\tilk_i\, \ck +\tilp_i \, \cp \\
 M_i^{V0,-+}({\bf p},{\bf k}) & =\tilk_i  \,\ck -\tilp_i \, \cp \\
 M_i^{V0,--}({\bf p},{\bf k}) & = 1 + \tilp_i \, \tilk_i \, \cp\, \, \ck \,  , \\
 \vspace{2mm}\nonumber\\
\Theta^{vj}_i=\gamma^j \quad (j=1,2,3):\hspace{-1cm} &\nonumber\\
M_i^{Vj,++}({\bf p},{\bf k}) & =-\tilp_i \,\cp \,\sigma_j-\tilk_i \, \sigma_j \, \ck \\
M_i^{Vj,+-}({\bf p},{\bf k}) & =-\sigma_j + \tilp_i \tilk_i \, \cp\, \sigma_j \,\ck \\
 M_i^{Vj,-+}({\bf p},{\bf k}) & =-\sigma_j + \tilp_i \tilk_i \, \cp\, \sigma_j \,\ck \\
 M_i^{Vj,--}({\bf p},{\bf k}) & =\tilp_i \cp \,\sigma_j+\tilk_i \sigma_j \ck \,.
\end{align}

 \end{appendix}

%%%%%%%%%%%%%%%%%%%%%% BIBLIOGRAPHY %%%%%%%%%%%%%%%%%%%%%%%%
\bibliographystyle{h-physrev3}
\bibliography{PapersDB-v2-4}

\end{document}